\documentclass[english,11pt,aps,prd,a4paper,preprintnumbers,floatfix,nofootinbib,showpacs,superscriptaddress, notitlepage]{revtex4-1}
 \pdfoutput=1

\usepackage[usenames,dvipsnames]{color}  
\usepackage{graphicx}
\usepackage{caption}
\usepackage{slashed}
\usepackage{subcaption}
\usepackage{amsmath}
\usepackage{amssymb}
\usepackage{bm}
\usepackage[colorlinks=true,citecolor=darkred,urlcolor=darkred, pdfborder={0 0 0}]{hyperref}
\usepackage[normalem]{ulem}
\usepackage{xcolor}

\definecolor{darkred}{rgb}{0.6,0,0}

\definecolor{linkcolor}{rgb}{0,0,0.5}

\def\gsim{\raise0.3ex\hbox{$\;>$\kern-0.75em\raise-1.1ex\hbox{$\sim\;$}}}
\def\lsim{\raise0.3ex\hbox{$\;<$\kern-0.75em\raise-1.1ex\hbox{$\sim\;$}}}
\def\beqn#1{\begin{equation}\label{#1}}
\def\eeqn{\end{equation}}
\def\beqa#1{\begin{eqnarray}\label{#1}}
\def\eeqa{\end{eqnarray}}

\def\Z2{$\mathcal{Z_2}$}

\def\vev#1{\left\langle #1\right\rangle}

\newcommand {\ignore}[1]{}

\def\321{$\mathrm{SU(3) \otimes SU(2) \otimes U(1)}$ }

 \def\SM{$\text{SU}(3)_c \otimes \text{SU}(2)_L \otimes \text{U}(1)_Y$ }

\newcommand{\AddrAHEP}{%
  AHEP Group, Institut de F\'{i}sica Corpuscular --
  CSIC/Universitat de Val\`{e}ncia, Parc Cient\'ific de Paterna.\\
 C/ Catedr\'atico Jos\'e Beltr\'an, 2 E-46980 Paterna (Valencia) - SPAIN}

\begin{document}

\bibliographystyle{unsrt}   

\title{ \color{BrickRed} 
\hspace {-3.5 mm} QCD axion, colour-mediated neutrino masses, and $B^+\to K^+ + E_{\text{miss}}$ anomaly\\
}

\author{Chandan Hati}
\email[Email Address: ]{chandan@ific.uv.es}
\affiliation{\AddrAHEP}
\author{Julio Leite}
\email[Email Address: ]{julio.leite@ific.uv.es}
\affiliation{\AddrAHEP}
\author{Newton Nath}
\email[Email Address: ]{Newton.Nath@ific.uv.es}
\affiliation{\AddrAHEP}
\author{Jos\'{e} W. F. Valle}
\email[Email Address: ]{valle@ific.uv.es}
\affiliation{\AddrAHEP}

\begin{abstract}

\vspace{1.0cm}
{\noindent
Motivated by the recent Belle II result indicating a $2.7\,\sigma$ excess of $B^+\to K^+ + E_{\text{miss}}$ events compared to the Standard Model (SM) prediction for $B^+ \to K^+ \nu\bar{\nu}$, we explore an explanation to this anomaly based on a KSVZ-type QCD axion model featuring a Peccei-Quinn (PQ) symmetry breaking at high-scale, that can provide a solution to the strong CP problem with dark matter relic abundance. The model contains a PQ-charged scalar} leptoquark which can interact with the SM quarks only via mass-mixing of the latter with vector-like quarks. The mixing between SM and vector-like quarks is determined by the PQ mass scales and can explain the excess $B^+\to K^+ + E_{\text{miss}}$ events while respecting other flavour constraints. The same PQ-charged scalar leptoquarks and vector-like quarks also mediate the two-loop radiative neutrino masses.

\end{abstract}
\maketitle



\section{Introduction}

It is well-known that within the Standard Model (SM), the rare semileptonic neutral-current meson decays are suppressed by a loop factor as well as by the Cabibbo-Kobayashi-Maskawa (CKM) suppression. Consequently, they provide a very clean probe for New Physics (NP) beyond the SM, particularly, if the final states are neutrinos. This is because, unlike in the case of the light-charged lepton in the final state, the modes involving neutrinos are not plagued by non-perturbative long-distance corrections~\cite{Altmannshofer:2009ma,Buras:2014fpa,Buras:2015yca,Blake:2016olu,Parrott:2022zte,Becirevic:2023aov}.

In its latest measurement, the Belle~II experiment has reported the branching ratio for the decay $B^+\to K^+ + E_{\text{miss}}$ to be $(2.3 \pm 0.5) \times 10^{-5}$, where $E_{\text{miss}}$ denotes missing energy. If one interprets the missing energy as a pair of light SM neutrinos, then this measurement presents a $2.7\, \sigma$ deviation from the SM prediction for $B\to K \nu \bar\nu$~\cite{Belle-II:2023esi}. The same quark-level transition $b \to s  \nu \bar \nu$ can also give rise to $B \to K^{\ast} +E_{\text{miss}}$, for which there is an upper limit from the BaBar experiment $\text{Br}(B^{ } \to K^{*}+E_{\text{miss}}) < 11 \times 10^{-5}$~\cite{BaBar:2013npw}. An upper limit also exists from the Belle experiment, which reported the combination of all $B \to K^* \nu\bar{\nu}$ channels  $\text{Br}(B \to K^* \nu\bar{\nu}) < 2.7 \times 10^{-5}$ at the 90\% CL~\cite{Belle:2017oht}, providing a tighter limit than the one from BaBar.  Another relevant limit that has recently been pointed out in Ref.~\cite{Alonso-Alvarez:2023mgc} is the branching fraction of the invisible $B_s$ decays, $\text{Br}(B_s \to E_{\rm miss}) < 5.4 \times 10^{-4}$ (90\% CL), by recasting the LEP data~\cite{ALEPH:2000vvi}. While the deviation could be due to a statistical fluctuation or some experimental misidentification, it is interesting to entertain the theoretical possibilities beyond the SM that can explain such a deviation.
Several studies have already explored these within an effective field theory approach, as well as using specific theoretical schemes~\cite{Bause:2023mfe,Allwicher:2023xba,Athron:2023hmz,Datta:2023iln,Chen:2024jlj,DAlise:2024qmp,Felkl:2023ayn,He:2023bnk,Berezhnoy:2023rxx,Altmannshofer:2023hkn,McKeen:2023uzo,Fridell:2023ssf,Ho:2024cwk,Gabrielli:2024wys,Hou:2024vyw,He:2024iju,Bolton:2024egx,Marzocca:2024hua,Rosauro-Alcaraz:2024mvx,Buras:2024ewl,Kim:2024tsm,Davoudiasl:2024cee,Greljo:2024evt}. A simple possibility is to assume heavy new physics contributions to $b \to s  \nu \bar \nu$ with the $\nu$ corresponding to the light SM neutrinos~\footnote{This can potentially be generalized to include a light right-handed neutrino exhibiting right-handed vector neutral current, in addition to the SM-like left-handed vector neutral current. Also, $\nu$ could be generalized to include some other light neutral fermion with mass less than $(m_{B^+}-m_{K^+})/2$.}. A minimal SM extension containing a scalar leptoquark with quantum numbers $(3,2,1/6)$, or a vector leptoquark with quantum number $(3,2,5/6)$ can do the job, potentially explaining the new measurement from Belle II consistent with the limit from BaBar~\cite{Bause:2023mfe,He:2023bnk}. 
Alternatively, the missing energy in the Belle II measurement can also be attributed to some other light new physics, such as a massive axion-like particle (ALP), $a$, exhibiting the $b \to s  a$ transition. Unfortunately, a true QCD axion does not lead to a very good fit for the new measurement. 

In this work, we consider the UV-complete KSVZ-type axion model developed in \cite{Batra:2023erw} to explore a potential connection between anomalous measurement and light neutrino mass generation.
This KSVZ-type model implements a global Peccei-Quinn (PQ) symmetry at high energy. After breaking the PQ symmetry around $10^{12}$ GeV, a light QCD axion arises, which can potentially explain the strong CP problem and the observed dark matter relic abundance. 
The model extends the SM with a new pair of vector-like isosinglet quarks, one scalar lepto-quark, and two coloured scalar mediators with non-zero PQ charges. This enables the possibility of generating flavour-changing currents and inducing light neutrino masses via two-loop quantum corrections. The anomalous Belle-II measurement can be explained in this scenario by a new contribution to $b \to s  \nu \bar \nu$ mediated by PQ-charged scalar leptoquarks also taking part in light neutrino mass generation, with the PQ-symmetry breaking scale tying everything neatly together. 

The plan for the rest of this paper is as follows. In section {\ref{EFT:anomaly}}, we discuss the model-independent vector-current explanations of the Belle-II $B^+\to K^+ + E_{\text{miss}}$ anomaly. In section \ref{sec:UVmodel}, we present our KSVZ-type axion model with the relevant particle content and interactions. 
In section \ref{sec:modelex}, we discuss the model proposed in section \ref{sec:UVmodel} in the context of the Belle-II $B^+\to K^+ + E_{\text{miss}}$ anomaly. We discuss the relevant new physics interaction flavour structure and the resulting constraints from the relevant flavour observables. In section \ref{sec:numass}, we discuss the two-loop neutrino mass generation mechanism in our model. In section \ref{sec:AxionPheno}, we examine the phenomenology associated with the QCD axion of the model. Finally, in section \ref{sec:ALP} we conclude, commenting on the alternative possibility of having an axion-like particle, instead of the true QCD axion, as the explanation of the Belle-II $B^+\to K^+ + E_{\text{miss}}$ anomaly.  

\section{Vector current explanations of the enhanced $B^+\to K^+ + E_{\text{miss}}$}{\label{EFT:anomaly}}

For the case of heavy new physics, we can parameterize the NP contributing to $b \to s  \nu \bar \nu$ through an effective interaction at around the $b$ mass scale. Such a theory is often referred to as the low energy effective field theory (LEFT) approach~\cite{Jenkins:2017jig,Liao:2020zyx}. 
We will consider the two dimension six vector operators that can contribute to $b\to s \nu\bar \nu$ given by~\footnote{ In the presence of light right-handed neutrinos there are two additional operators with right-handed vector currents between the neutrino pairs~\cite{He:2023bnk}. The light neutrinos can also be generalized to some heavy long-lived neutral fermion producing a missing energy signature at Belle-II, see Refs.~\cite{Fridell:2023ssf,Bolton:2024egx} for some relevant discussion and global fits.}
\begin{equation}
\label{LEFTvector}
    \mathcal{L}_{\rm LEFT}^{bs\nu\nu} \supset \frac{8 G_F V_{tb} V_{ts}^*}{\sqrt{2}} \frac{\alpha}{4\pi}
    \sum_{\alpha\beta}\left[ (C_{L}^{sb\alpha\beta} + C_{L}^{sb, \rm{ SM}} \delta^{\alpha \beta} ) 
   (\bar{s}_{L}\gamma_{\mu} b_{L}) (\bar\nu_L^\alpha \gamma^\mu \nu_L^\beta) +
   C_{R}^{sb\alpha\beta} (\bar{s}_{R}\gamma_{\mu} b_{R}) (\bar\nu_L^\alpha \gamma^\mu \nu_L^\beta) \right] \ ,
\end{equation}
where $C_L^{sb, \rm SM} \approx -6.32$  corresponds to the contribution present in the SM~\cite{Buchalla:1998ba,Brod:2010hi,Becirevic:2023aov}. The indices $\alpha,\beta$ in the superscript correspond to the SM neutrino flavours. To discuss global fits, it is convenient to define the ratios of the measured branching fractions with respect to the SM contribution
\begin{equation}
R_{K^{(*)}}^\nu=\frac{\text{Br}(B \to K^{(*)} \nu\bar{\nu})}{\text{Br}(B \to K^{(*)} \nu\bar{\nu})_{\mathrm{SM}}} \ . 
\end{equation}
The latest Belle II measurement of $B^+\to K^+ \nu \bar\nu$ announced a value of $R^\nu_{K} = 5.3 \pm 1.7$~\cite{Belle-II:2023esi}.
Moreover, the new measurement from Belle-II, when combined with the previous upper limit on the same mode, has been shown to imply $R^\nu_{K} = 2.93 \pm 0.90$~\cite{Marzocca:2024hua}, which corresponds to $2.1\,\sigma$ deviation from the SM.
On the other hand, the tighter limits from the Belle experiment on the combination of all $B \to K^* \nu\bar{\nu}$ channels~\cite{Belle:2017oht} translate to $R^\nu_{K^*} = 1.0 \pm 1.1$ ~\cite{Marzocca:2024hua}, assuming the central value as given by the SM prediction. In terms of the Wilson coefficients in Eq. (\ref{LEFTvector}), the ratios $R^\nu_{K}$ and $R^\nu_{K^{*}}$ can be analytically expressed as~\cite{Browder:2021hbl,He:2023bnk}
\begin{subequations}
\begin{align}
R_K^{\nu}  = & 1 
+ {2 C_{L}^{sb,\rm{SM}} \over 3 |C_{L}^{\rm{SM}}|^2}
\sum_\alpha {\rm Re} ( C^{sb\alpha\alpha}_{L}+C^{sb\alpha\alpha}_R )
+{1\over 3 |C_{L}^{sb,\rm{SM}}|^2 } \sum_{\alpha\beta}\left|C^{sb\alpha\beta}_L+C^{sb\alpha\beta}_R\right|^2 ,
\\
R_{K^*}^{\nu}  = & 1 
+ {2 C_{L}^{sb,\rm{SM}} \over 3 |C_{L}^{sb,\rm{SM}}|^2}
\sum_\alpha {\rm Re} \left( C^{sb\alpha\alpha}_{L} \right)
+{1\over 3 |C_{L}^{sb,\rm{SM}}|^2 } 
\sum_{\alpha\beta}\left( \left|C^{sb\alpha\beta}_L\right|^2 +\left|C^{sb\alpha\beta}_R\right|^2 \right)
\nonumber
\\
& - 2 \eta \left[
{C_{L}^{sb,\rm{SM}} \over 3 |C_{L}^{sb,\rm{SM}}|^2}
\sum_\alpha {\rm Re} \left( C^{sb\alpha\alpha}_{R} \right)
+{1\over 3 |C_{L}^{sb,\rm{SM}}|^2 } 
\sum_{\alpha\beta} {\rm Re}\left(C^{sb\alpha\beta}_LC^{sb\alpha\beta}_R\right)
\right], 
\label{eq:scavecbr}
\end{align}
\end{subequations}
where 
\begin{align}
  \eta \equiv {F_- \over F_+}, \,
  F_\pm \equiv \int_0^{(1 -\sqrt{x_K})^2} dx 
  \lambda^{1\over 2}(1, x_K, x) \left[ x A_1^2 + 
  {32\, x_K \over (1 + \sqrt{x_K})^2} A_{12}^2
  \pm  { x \lambda(1, x_K, x) \over (1 + \sqrt{x_K})^4 } V_0^2  \right], 
\end{align}
with $x \equiv q^2/m_B^2$, $x_K \equiv m_{K^*}^2/m_B^2$ and $\lambda(a,b,c)=[a^2-(b+c)^2] [a^2-(b-c)^2]$.
Numerically, $\eta = 0.63\pm 0.09$ when using the latest light-cone sum rule (LCSR) result for the form factors $A_{1}, A_{12}, V_0$~ \cite{Gubernari:2018wyi},
${2 C_{L}^{sb,\rm{SM}} / 3 |C_{L}^{sb,\rm{SM}}|^2} = -(0.105\pm 0.001)$ and ${1 / 3 |C_{L}^{sb,\rm{SM}}|^2} = 0.0083\pm 0.0002$. Note that the new contributions can be lepton flavour universal or prefer some particular neutrino flavour (e.g. $\tau$). To this end, it has been noted (see e.g. Refs.~\cite{Bolton:2024egx}) that the significant new contribution needed to explain the deviation from the SM using the operators in Eq. (\ref{LEFTvector}), will also enhance the $b\to s\ell\ell$ transition, with $\ell$ denoting the charged lepton generations. The latter is subject to constraints from the measurements of $B\to K^{(*)} \ell\ell$ and $B_s \to \ell^+\ell^-$. 

Given the tight constraints on the two lightest leptons, one of the viable phenomenological options is to assume that the NP couples mainly to $\tau$ neutrinos, which is relatively loosely contained. In what follows, we will focus on such lepton flavour universality violating scenario. The decay rate for the invisible decay of $B_s$ to a pair of nearly massless SM neutrinos is given by 
\begin{align}{\label{Bsinv}}
\Gamma(B_s \to \nu_\alpha\bar{\nu}_\alpha) =  \frac{f_{B_s}^2 m_{B_s}^3 C_N^2}{32\pi} \sqrt{1 - \frac{4m_{\nu_\alpha}^2}{m_{B_s}^2}}\left(|C^{sb\alpha\alpha}_{L,\rm{tot}} -C^{sb\alpha\alpha}_R|^2\frac{m_{\nu_\alpha}^2(m_b+m_s)^2}{m_{B_s}^4} \right)\frac{m_{B_s}^2}{(m_b+m_s)^2}\,,
\end{align}
where $C_N=\frac{8 G_F V_{tb} V_{ts}^*}{\sqrt{2}} \frac{\alpha}{4\pi}$, $f_{B_s}=230.3 (1.3)$ MeV denotes to the $B_s$ decay constant, and $C^{sb\alpha\alpha}_{L,\rm{tot}}=C^{sb,SM}_{L}+C^{sb\alpha\alpha}_{L}$. From Eq. (\ref{Bsinv}), one sees that the decay rates are highly suppressed by the light neutrino mass squared. We note that the constraint $\text{Br}(B_s \to E_{\rm miss}) < 5.4 \times 10^{-4}$ (90\% CL) allows for the whole parameter space.
\begin{figure}[t!]
		\centering
		\includegraphics[width=0.6 \textwidth]{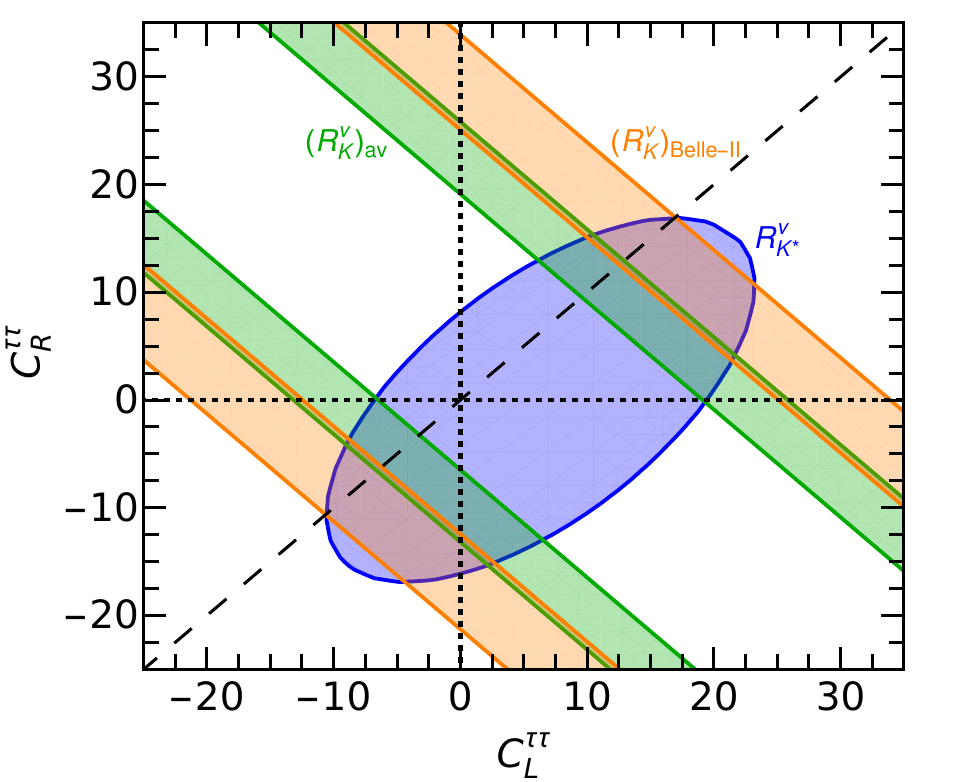} 
		\caption{\footnotesize Region in the  $(C^{\tau\tau}_{L}-C^{\tau\tau}_{R})$ plane, allowed by the favoured $R^\nu_{K^+}$, $R^\nu_{K^*}$ values. The blue contour delineates the allowed $R^\nu_{K^*}$ parameter space, and the green and orange contours represent the regions consistent with the average value of $R^\nu_{K^+}$ and the latest measurement from Belle II, respectively. Dotted black lines correspond to $C^{\tau\tau}_{L} = 0$ and $C^{\tau\tau}_{R} = 0 $, respectively, whereas the diagonal dashed line depicts $C^{\tau\tau}_{L} = C^{\tau\tau}_{R}$ (see text for details). We note that our bands correspond to the uncertainties quoted by the relevant experimental measurements (or average of them) with the errors added in quadrature, which can approximately be considered as $1\sigma$ uncertainty bands.}
		\label{fig:CLCRplot} 
	\end{figure}

In Fig. \ref{fig:CLCRplot}, we show allowed regions in the $C^{sb\tau\tau}_{L} (\equiv C^{\tau\tau}_{L})$ vs $C^{sb\tau\tau}_{R} (\equiv C^{\tau\tau}_{R})$ plane describing the areas consistent with preferred values of $R^\nu_{K^+}$, $R^\nu_{K^*}$ and the constraints from $B_s$ invisible decays, under the conservative assumption that $R^\nu_{K^{+}}$ dominates $R^\nu_{K^*}$.
From the plot, it is clear that only  $C^{\tau\tau}_{L}$ cannot by itself explain the anomalous data while being consistent with the constraint from the Belle experiment on the combination of all $B \to K^* \nu\bar{\nu}$ channels. 

In contrast,  $C^{\tau\tau}_{R}$ can successfully do so for $C^{\tau\tau}_{R} \in [-6.49, -13.19]$ with $C^{\tau\tau}_{L} = 0$ as shown by the green regions. 
We also find that assuming only the Belle II measurement restricts $C^{\tau\tau}_{R}  \in [-12.41, -16.20]$ with $C^{\tau\tau}_{L} = 0$, see the overlap regions between blue and orange contours.
In the following section, we will propose a model involving a scalar with quantum number $(3,2,1/6)$ and a PQ charge that will precisely generate only $C^{\tau\tau}_{R}$.
\section{A KSVZ-type QCD Axion Model}\label{sec:UVmodel}
The model we investigate here is inspired by the framework proposed in Ref. \cite{Batra:2023erw}, in which exotic fermions mix with the SM down-type quarks, as discussed in the supplemental material of that paper. 
Besides the SM content, the model includes the states listed in the right panel of Table~\ref{tab:general}. Multiplicities are indicated in the last column of each panel. 

Notice that the model features a global Peccei-Quinn symmetry, $U(1)_{PQ}$, under which all the new fields, except for $\Psi_L$, are charged. Moreover, apart from $\sigma$ and $\Sigma$, which are singlet and anti-sextet, respectively, all fields are triplets under $SU(3)_C$. As for the SM fields, they are not charged under $U(1)_{PQ}$. 
As we will see, this model inherently gives rise to new contributions to the $b\to s \nu\bar \nu$ process.
\begin{table}[!h]
    \centering
\begin{tabular}{ cc }  
	\begin{tabular}{ p{1.5cm}  c  c   c  }
		\hline
 Fields & $\mathcal{SM}$ &    $~~~U(1)_{PQ}~~~~$ & Families\\
		\hline 
\quad$L_L$&  $ (1, 2, -1/2)$  & $0$  & $3$\\
\quad$e_R$&  $(1, 1, -1) $  &  $0$ & $3$ \\
\quad$Q_L$&  $ (3, 2, 1/6)$  & $0$  & $3$\\
\quad$d_R$&  $(3, 1, -1/3) $  &  $0$ & $3$ \\
\quad$u_R$&  $(3, 1, 2/3) $  &  $0$ & $3$ \\
  \hline 
\quad$\Phi$ &($1,2, 1/2$)& $0$ & 1\\
\hline
	\end{tabular}
\hspace{1cm}
\begin{tabular}{ p{1.5cm}  c  c   c  }
		\hline
 Fields & $\mathcal{SM}$ &    $~~~U(1)_{PQ}~~~~$ & Multiplicities\\
		\hline 
\quad$\Psi_L$&  $ (3, 1, -1/3)$  & $0$  & $2$\\
\quad$\Psi_R$&  $(3, 1, -1/3) $  &  $-1/2$ & $2$ \\
		\hline 
\quad$\sigma$ &($1,1, 0$)& $1/2$ & 1\\
\quad$\eta$ & $(3, 2, 1/6)$ & $-1/2$ & $1$\\
\quad$\chi$ & ($3, 1, -1/3$)& $-1/2$ & $1$ \\
\quad$\Sigma$ & ($\bar{6}, 1, 2/3$)& $1$ & $1$ \\
\hline
\end{tabular} 
\end{tabular}
\caption{ \footnotesize The left and right panels list the Standard Model and new fields and their quantum numbers, where $\mathcal{SM} \equiv$ \SM. Except for $\Psi_L$, all the new fields carry a PQ charge.}
\label{tab:general} 
\end{table}

The Yukawa interactions relevant for explaining the Belle-II anomaly and generating neutrino masses involve the new fields and the usual SM down-type Yukawa terms
\begin{align}
		- \mathcal{L}_{\text{Yuk.}} \supset Y_{ab}^{\Psi} \overline{\Psi_{aL}} \Psi_{bR} \sigma + \frac{Y^{\Sigma}_{ab}}{2}  {\Psi^T_{aR}}\,C\,\Sigma\,\Psi_{bR} + Y_{ia} \overline{L_{iL}}\, \tilde{\eta} \, \Psi_{aR} + M^{\Psi d}_{ai} \,\overline{\Psi_{aL}} d_{iR}  + y^d_{ij}\, \overline{Q_{iL}} \,\Phi\,d_{jR} + \text{H.c.}\,,
		\label{eq:Yuk}
	\end{align}
where $\tilde{\eta} = i \sigma_2 \eta^\ast $, and $Q_{iL}$, $d_{jR}$ and $\Phi$ are the left-handed SM quark doublets, right-handed down-type quarks and the Higgs doublet, respectively, with $a,b=1,2$ and $i,j=1,2,3$. $C$ stands for the charge conjugation operator. Furthermore, the most general scalar potential can be written as
\begin{eqnarray}\label{eq:V}
\mathcal{V}\hspace{-0.2cm}&=&\hspace{-0.2cm}
\sum_S \left[ \mu_{S}^2S^{\dag}S+\frac{\lambda_S}{4}\left(S^{\dag}S\right)^2\right] + \sum_{S\neq S^\prime}
\frac{\lambda_{SS^\prime}}{2}\left(S^{\dag}S\right)\left({S^\prime}^{\dag}S^\prime\right)\nonumber\\
&&\!\!\!
+\mu^2_{\Sigma }\, \mbox{Tr} \left( \Sigma^\dagger \Sigma \right) 
+\lambda_{\Sigma }\, \mbox{Tr} \left( \Sigma^\dagger \Sigma \right)^2 + \lambda_{\Sigma^\prime}\,\left(\mbox{Tr}\, (\Sigma^\dagger \Sigma ) \right)^2 
+  \sum_{S}
\frac{\lambda_{S \Sigma}}{2}\left(S^{\dag}S\right)\mbox{Tr}\left({\Sigma}^{\dag}\Sigma\right)
\nonumber\\
&&\!\!\!
+\tilde{\lambda}_{\Phi\eta}\left(\Phi^{\dag}\eta\right)\left(\eta^{\dag}\Phi\right) + \left( \mu \,\chi\, \Sigma \,\chi 
+ \kappa\, \eta^\dagger \Phi \, \chi + \lambda\, \Phi^\dagger\, \eta\, \Sigma \,\chi 
+\mathrm{h.c.}\right),
\end{eqnarray}
where $S,S^\prime = \Phi, \eta, \chi, \sigma$. The last three terms, as we discuss below, are relevant for neutrino mass generation. 

We assume that $\sigma$ acquires a vacuum expectation value (vev), spontaneously breaking $U(1)_{PQ}$ at a large energy scale, i.e.
$\vev{\sigma} = v_\sigma/\sqrt{2} \simeq f_a$, with $f_a$ being the axion decay constant.
On the other hand, the Standard Model Higgs doublet, $\Phi = (\phi^+\,\phi^0)^T$, breaks the electro-weak symmetry via its vev $\langle \phi_0\rangle = v/\sqrt{2}$, with $v=246$ GeV. The remaining scalar fields, $\eta,\chi$, and $\Sigma$ carry colour and electric charges and, as such, must not acquire vevs.

Once $U(1)_{PQ}$ and electroweak spontaneous symmetry breaking take place, one defines a $5 \times 5$ mass matrix for the extended down-type sector from $\overline{{\bf B}_L} \, \bm{\mathcal{M}}^B \,{\bf B}_R$, in the basis ${\bf B}_{L,R} =(\bm{d}\,,\,\bm{\Psi})_{L,R}$, 
\begin{equation}
\bm{\mathcal{ M}}^B =
    \begin{pmatrix}
    \bm{m}^d_{[3\times3]} & \bm{0}_{[3\times 2]} \\
    \bm{M}^{\Psi d}_{[2\times3]} &  \bm{M}^\Psi_{[2\times2]}
    \end{pmatrix}\,,
    \label{eq:mixmat}
\end{equation}
where $m^d_{ij} =y^d_{ij} v/\sqrt{2}$, $M^\Psi_{ab} =Y^{\Psi}_{ab} v_\sigma/\sqrt{2}$, with $i,j=1,2,3$ and $a,b=1,2$. Note that, hereafter, we denote matrices by bold characters, while their elements are in regular characters unless specified otherwise.  The numbers in square brackets represent the dimensions of the sub-matrices.

Assuming the seesaw limit $m^d\ll M^{\Psi d}\,, M^\Psi$, we can block diagonalise the $\bm{\mathcal{M}}^B{\bm{\mathcal{M}}^{B}}^\dagger$ to obtain 
\begin{align}
\bm{\Lambda}^{d}_{[3\times 3]} & \simeq \bm{m}^d
\left(\bm{\mathcal{I}}_3-{\bm{M}^{\Psi d}}^{\dagger}{\bm{\Lambda}^{\Psi}}^{-1} \bm{M}^{\Psi d}\right){\bm{m}^d}^\dagger\,,\\
    \bm{\Lambda}^{\Psi}_{[2\times 2]} & \simeq \bm{M}^\Psi {\bm{M}^\Psi}^\dagger + \bm{M}^{\Psi d} {\bm{M}^{\Psi d}}^\dagger   
\,.\nonumber
\end{align}
Furthermore, for $M^{\Psi d}\ll M^\Psi$, one finds that $\Lambda^d$ becomes
\begin{align}
    \bm{\Lambda}^{d}_{[3\times 3]} & \simeq \bm{m}^d{\bm{m}^d}^\dagger - \bm{m}^d
{\bm{M}^{\Psi d}}^{\dagger}(\bm{M}^\Psi {\bm{M}^\Psi}^\dagger)^{-1} \bm{M}^{\Psi d}{\bm{m}^d}^\dagger
\,.
\end{align}
Therefore, in this limit, the leading contributions to light quark masses come mainly from the Yukawa interactions with the SM Higgs field, whereas contributions due to the mixing with the heavy vector-like quarks are sub-leading. Finally, the matrix $\bm{\Lambda}^d$ is diagonalized by a $3 \times 3$ unitary matrix, $\bm{V}^d_L$, which is closely related to the CKM matrix~\footnote{Here $\bm{V}_{CKM} = \bm{V}^{u\,\dagger}_L \bm{V}^d_L$, where $\bm{V}^{u}_L$ is the $3\times 3$ unitary matrix that diagonalises $\bm{m}^u{\bm{m}^u}^\dagger$, with $\bm{m}^u$ denoting the up-type quark mass matrix.}, while $\bm{\Lambda}^\Psi$ is diagonalized by $\bm{V}^\Psi_L$, a $2 \times 2$ unitary matrix in the heavy sector.

The diagonalization above was achieved by performing appropriate unitary transformations on the left-hand fields. Similarly, we can diagonalize ${\bm{\mathcal{M}}^{B}}^\dagger \bm{\mathcal{M}}^B$ by rotating the right-handed fields. 
In general, we can parametrise the transformations relating the symmetry and the mass bases via
\begin{align}
    \begin{pmatrix}
    \bm{d} \\
    \bm{\Psi}
    \end{pmatrix}_{L,R}
    =  \begin{pmatrix}
    (\bm{\mathcal{I}}_{3}- \bm{\Theta} \bm{\Theta}^\dagger)^{1/2} & \bm{\Theta} \\
    -\bm{\Theta}^\dagger & (\bm{\mathcal{I}}_{2}-  \bm{\Theta}^\dagger \bm{\Theta})^{1/2}
    \end{pmatrix}_{L,R}
    \begin{pmatrix}
    \bm{V}^d & 0 \\
    0 & \bm{V}^\Psi
    \end{pmatrix}_{L,R}
    \begin{pmatrix}
    \bm{d}^\prime \\
   \bm{\Psi}^\prime
    \end{pmatrix}_{L,R},
    \label{eq:qtrans}
\end{align}
with the primed fields representing the mass eigenstates, and $\bm{\mathcal{I}}_{2},\bm{\mathcal{I}}_{3}$ are $2\times 2$ and $3\times 3$ identity matrices, respectively. Specifically, the mixing patterns amongst light and heavy quarks are given by the $3\times 2$ matrices that follow
\begin{align}\label{eq:qmix} 
    \bm{\Theta}_R &\simeq {\bm{M}^{\Psi d}}^\dagger\,{{\bm{M}^\Psi}^\dagger}^{-1}\, ,\\
    \bm{\Theta}_L &\simeq \bm{m}^d\,{\bm{M}^{\Psi d}}^\dagger\,{\bm{\Lambda}^\Psi}^{-1}\simeq \bm{m}^d\,\bm{\Theta}_R\, {\bm{M}^\Psi}^{-1}\,. \nonumber
\end{align} 
Hence $\bm{\Theta}_L\ll \bm{\Theta}_R \ll 1$ in such a way that we can safely neglect contributions coming from $\bm{\Theta}_L$.

\section{Explaining the enhanced $B^+\to K^+ + E_{\text{miss}}$ rate}
\begin{figure}[t!]
		\centering
         \includegraphics[scale=0.3]{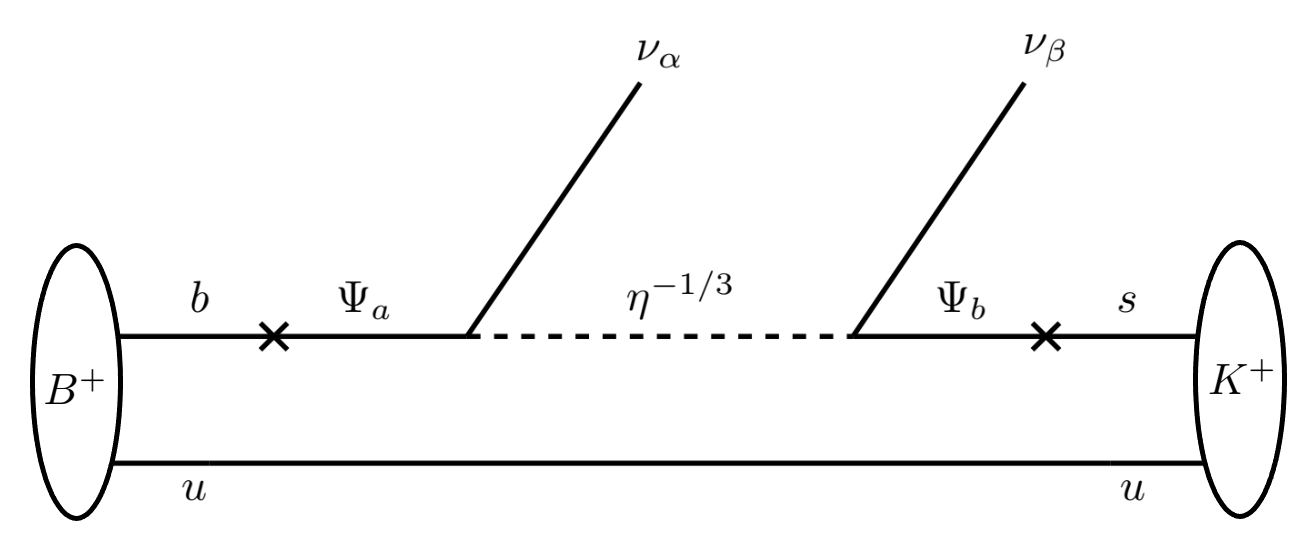} \hspace{+0.1cm}
		\caption{ Leptoquark-mediated $b \to s \nu\nu $ induced by heavy-standard quark mixing.}
		\label{fig:B2Knunu}
	\end{figure}

{\label{sec:modelex}}
As a consequence of the mixing between SM and the vector-like quarks, parameterised in terms of $\bm{\Theta}_L$ and $\bm{\Theta}_R$, flavour-changing interactions arise in the down-type quark sector.
They are mediated not only by the SM $Z$ boson and Higgs field $h$, but also by other fields such as the axion $a$, the $\eta$ and
the $\Sigma$~\footnote{In Appendix \ref{app:FCNCs}, we derive the quark flavour-changing neutral currents mediated by $Z$, $h$ and $a$.}. The flavour-changing interactions induced by $\eta$ couplings give an additional contribution to $B^+\to K^+ + E_{\text{miss}}$. The field $\eta$ has the same quantum numbers as a standard doublet leptoquark. However, it carries a nontrivial PQ charge, which prevents a direct coupling of $\eta$ with SM quarks. Instead, this interaction must occur through the mixing of SM quarks with the vector-like quarks.
In the mass basis for quarks and charged leptons,  the couplings of  $\eta = (\eta^{-2/3},\, \eta^{1/3})^T $ in Eq.~\eqref{eq:Yuk} take the form 
\begin{eqnarray}
    \mathcal{L}_{\eta}^{\text{mass}} =   \overline{\nu_{\alpha L}}\, {{\lambda}}_{\alpha \beta}  \, d_{\beta R}\,\eta^{1/3}- \overline{\ell_{\alpha L}}\, {{\lambda}'}_{\alpha\beta}   \, d_{\beta R}\,\eta^{-2/3} +\text{h.c.} \, ,
    \label{eq:etamb}
\end{eqnarray}
where
\begin{align}\label{eq:coupmb}
  [{\bm{\lambda}}]_{\alpha \beta} &= -   [{\bm{Y}}]_{\alpha a} [{\bm{\Theta}^\dagger_R}]_{ap} [{\bm{V}^{du}_R}]_{p\beta} \, ,\\
  [{\bm{\lambda}'}]_{\alpha \beta} &= - [{{\bm{U}_L^{\nu \ell}}^\dagger}]_{\alpha i}[\bm{Y}]_{i a} [{\bm{\Theta}^\dagger_R}]_{ap} [{\bm{V}^{du}_R}]_{p\beta}\, , \nonumber
\end{align}
where we follow the unconventional but convenient choice that the lepton mixing matrix $\bm{U}_L^{\nu \ell}$, diagonalising the standard left-handed leptonic charged current, is attached to the charged leptons. Note that we follow the convention of bold characters enclosed in square brackets when the identities hold both at the matrix and specified matrix element level, with repeated dummy indices summed over, in Einstein's summation convention. To simplify the analysis we choose to work on a basis where in Eq.~\eqref{eq:etamb} the down-type quarks are diagonal up to the mixing with vector-like quarks, implying that  $\bm{V}^{du}_R\equiv \bm{I}_{3\times 3}$. This simplification holds unless a right-handed charged current exists, making the right-handed analogue of the CKM matrix physically observable and $\eta$ interactions are not diagonal in the diagonalised right-handed charged current basis. Under the above simplification Eq.~\eqref{eq:coupmb} reduces to
\begin{align}\label{eq:coupmb1}
   [{\bm{\lambda}}]_{\alpha \beta} &= -   [\bm{Y}]_{\alpha a} [{\bm{\Theta}^\dagger_R}]_{a\beta} \, , \\
   [{\bm{\lambda}'}]_{\alpha \beta} &= - [{{\bm{U}_L^{\nu \ell}}^\dagger}]_{\alpha i}[\bm{Y}]_{i a} [{\bm{\Theta}^\dagger_R}]_{a\beta}\,. \nonumber
\end{align}
A small non-unitarity in the total SM down-type quark diagonalising matrix arises due to the presence of the additional vector-like quark block~\cite{Schechter:1981cv,Xing:2007zj,Escrihuela:2015wra,Blennow:2016jkn,Fernandez-Martinez:2016lgt,Hati:2019ufv}, which we will also neglect for simplicity. The relevant Lagrangian due to $\eta$-mediated FCNC interaction $b\to s \nu_{\tau} \overline{\nu_{\tau}}$ transition can then be written using   Eq.~\eqref{eq:etamb}  as
\begin{eqnarray}
    \mathcal{L}_{\eta}^{\text{mass}} =   \overline{\nu_{\tau L}}\, {{\lambda}}_{3 3} \, b_{R}\,\eta^{1/3} +
      \overline{\nu_{\tau L}}\, {{\lambda}}_{32} \, s_{R}\,\eta^{1/3} + 
    \overline{b_R}\, {{\lambda}}_{33}^* \, \nu_{\tau L}\,\eta^{-1/3}\, +
  \overline{s_R}\, {{\lambda}}_{32}^* \, \nu_{\tau L}\,\eta^{-1/3}\,.
    \label{eq:etamb1}
\end{eqnarray}
The corresponding Feynman diagram for the $\eta$-mediated $b\to s \nu_{\tau}\overline{\nu_{\tau}}$ transition is shown in Fig.~\ref{fig:B2Knunu}.

The effective Lagrangian can be expressed, using the first and last terms of Eq.~(\ref{eq:etamb1}), as
\begin{equation}
\label{eq:Lbsnunu-model}
    \mathcal{L}_{\rm eff}^{bs\nu\nu} \supset  -
   \frac{{{\lambda}^*_{32}} {\lambda_{33}}}{2 m_\eta^2} \, (\bar{s}_{R}\gamma_{\mu} b_{R}) \, (\bar\nu_{\tau L} \gamma^\mu \nu_{\tau L}) \,,
\end{equation}
where we assume that the field $\eta$ is much heavier than the scale of the FCNC $B$-meson decay process and therefore can be integrated out. Matching this interaction with the effective interactions in Eq. (\ref{LEFTvector}), the relevant Wilson coefficient of interest for $\alpha=\alpha'=3$ (i.e. the neutrinos are of tau flavour) is then given by
\begin{equation}
\label{eq:ch:CR-model}
C_{R}^{sb\tau \tau}  = -\left(\frac{8 G_F V_{tb} V_{ts}^*}{\sqrt{2}}\frac{\alpha}{4\pi}\right)^{-1}  
   \frac{{\lambda_{32}^*} {\lambda_{33}}}{2 m_\eta^2}\,.
\end{equation}
%

\subsection{ Flavour structure of scalar leptoquark and vector-like quark couplings}

To explore the feasibility of explaining the combined fit to the enhanced $B^+\to K^+ + E_{\text{miss}}$ branching fractions discussed in Section \ref{EFT:anomaly} in terms of our proposed theory, we now make some simplifying assumptions.
Without loss of generality, we assume the heavy vector-like states to be diagonal in Eq.~\eqref{eq:mixmat},
\begin{equation}\label{eq:mfex1}
    \bm{M}^{\Psi }=
\begin{pmatrix}
   {M^{\Psi}_{1}}  & 0   \\
0 & {M^{\Psi }_{2}}   
    \end{pmatrix}\,.
\end{equation}
On the other hand, the most general mixing matrix connecting the  heavy vector-like and the SM down quark states can be expressed as
\begin{equation}\label{eq:mfex2}
 \bm{M}^{\Psi d}=
\begin{pmatrix}
   {M^{\Psi d}_{11}}  & {M^{\Psi d}_{12}}  & {M^{\Psi d}_{13}} \\
 {M^{\Psi d}_{21}} & {M^{\Psi d}_{22}}  & {M^{\Psi d}_{23}} 
    \end{pmatrix}\,.
\end{equation}
Utilizing Eq.~\eqref{eq:mfex1} and Eq.~\eqref{eq:mfex2}  in Eq.~\eqref{eq:qmix}, one finds
\begin{equation}\label{eq:mfex3}
\bm{\Theta}_R^{\dagger}=\begin{pmatrix}
    \frac {M^{\Psi d}_{11}} {M^{\Psi}_{1}} & \frac{M^{\Psi d}_{12}} {M^{\Psi}_{1}} & \frac{M^{\Psi d}_{13}} {M^{\Psi}_{1}}\\
     \frac {M^{\Psi d}_{21}} {M^{\Psi}_{2}} & \frac{M^{\Psi d}_{22}} {M^{\Psi}_{2}} & \frac{M^{\Psi d}_{23}} {M^{\Psi}_{2}}
    \end{pmatrix} \, .
\end{equation}
Since we are primarily interested in the scenario where the PQ-charged scalar leptoquark dominantly couples to the two heavier generations of leptons and down-type quarks, we can take $Y_{i a}$ in Eq. \eqref{eq:Yuk} to be
\begin{equation}\label{eq:mfex4}
 \bm{Y}=\begin{pmatrix}
    0 & 0\\
      Y_{21} & Y_{22} \\
     Y_{31} & Y_{32}
    \end{pmatrix} \, .
\end{equation}
Since the couplings to the first generation of charged leptons and quarks are subject to much tighter constraints from charged lepton flavour violating (cLFV) processes and meson decays involving the lighter leptons, they should be suppressed in order to be consistent with the experiment. For instance, in the presence of sizeable couplings of $\eta$ to both electron and muons, then cLFV processes like $\mu \to e \gamma$, $\mu \to 3e$, $\mu-e$ conversion in nuclei would be induced via one-loop diagrams involving heavy vector-like and SM quark fields in the loop.
Such cLFV processes provide too stringent constraints on flavour violating leptoquark couplings~\cite{Dorsner:2016wpm,Becirevic:2016yqi,Hati:2018fzc,Angelescu:2018tyl,Hati:2019ufv,Crivellin:2021bkd,Fajfer:2024uut} to contribute sizeably to $b\to s \nu\overline{\nu}$ rate so as to explain the Belle -II excess. Therefore, in this paper we assume that any significant New physics contribution to arise in tau flavor and assume the couplings to electron and muons to be SM like. We note that allowing for substantial coupling to muons is subject to strong constraints from the $b \rightarrow s \mu \mu $ data, where the universality ratios are now in very good agreement with the SM~\cite{Gubernari:2022hxn,Ciuchini:2022wbq,Alguero:2023jeh,Wen:2023pfq}.  

There are also constraints from atomic parity violation~\cite{Crivellin:2021bkd}, parity-violating deep inelastic scattering~\cite{Crivellin:2021bkd}, and coherent elastic neutrino-nucleus scattering~\cite{DeRomeri:2023cjt} on the first-generation couplings.
 Moreover, the constraints from the experimental measurements of $K \to \pi \nu\nu$ are more stringent than for mesons involving heavier quarks, see e.g.~\cite{Dorsner:2016wpm,Bobeth:2017ecx,Mandal:2019gff,Fridell:2023rtr,Buras:2024ewl,Crivellin:2021bkd} for some relevant discussion. The $K^+ \to \pi^+ \nu \bar{\nu}$ and $K_L \to \pi^0 \nu \bar{\nu}$ modes are currently under investigation by the NA62 and KOTO Collaborations, respectively. NA62 has found a $3.4\sigma$ evidence for the charged mode with $\text{Br}(K^+ \to \pi^+ \nu\bar{\nu}) = (10.6^{+4.0}_{-3.4} \pm 0.9 ) \times 10^{-11}$. A final sensitivity of about 15\% is expected by NA62, while KOTO should reach a 95\%CL upper limit of $1.8 \times 10^{-10}$. In addition, there are also constraints from $K^0-\bar{K}^0$ mixing~\cite{Brod:2011ty}, $K_L\to  \mu e$, $K_L\to  \pi \mu e$~\cite{Plakias:2023esq}, etc. Using Eqs. \eqref{eq:mfex3} and \eqref{eq:mfex4} in Eq.~\eqref{eq:coupmb1} and neglecting first generation mixing terms for simplicity, i.e., $M^{\Psi d}_{11}\sim M^{\Psi d}_{21} \to  0$, we get
\begin{equation}\label{eq:mfex6}
\bm{\lambda}=\begin{pmatrix}
    0 & 0 & 0\\
    0 & Y_{21} \frac {M^{\Psi d}_{12}} {M^{\Psi}_{1}}+Y_{22} \frac {M^{\Psi d}_{22}} {M^{\Psi}_{2}} & Y_{21} \frac {M^{\Psi d}_{13}} {M^{\Psi}_{1}}+Y_{22} \frac {M^{\Psi d}_{23}} {M^{\Psi}_{2}}\\
     0 & Y_{31} \frac {M^{\Psi d}_{12}} {M^{\Psi}_{1}}+Y_{32} \frac {M^{\Psi d}_{22}} {M^{\Psi}_{2}} & Y_{31} \frac {M^{\Psi d}_{13}} {M^{\Psi}_{1}}+Y_{32} \frac {M^{\Psi d}_{23}} {M^{\Psi}_{2}}
    \end{pmatrix} \, ,
\end{equation}
where we recall that in the element $\lambda_{ij}$ the first index corresponds to the neutrino flavour and the second one to the quark flavour, c.f. Eq.~(\ref{eq:etamb}).

\subsection{Explanation of $B^+\to K^+ + E_{\text{miss}}$ anomaly and relevant constraints}

The minimal couplings required to generate $b\to s \nu_{\tau}\overline{\nu_{\tau}}$ are $\lambda_{32}$ and $\lambda_{33}$. This can be achieved in our model if $Y_{3i}$, $\frac{M^{\Psi d}_{i2}}{M^{\Psi}_{i}}$ and $\frac{M^{\Psi d}_{i3}}{M^{\Psi}_{i}}$ are nonzero for $i=1$ or 2. We would also like to note that the elements of the matrix $\bm{\lambda}$, describing the effective scalar leptoquark $\eta$-couplings with the left-handed SM lepton doublet and right-handed down-type quark, correspond to those of the standard doublet leptoquark in the literature. The crucial point to highlight here is that the effective couplings are naturally suppressed in our model by the PQ scale due to the mixing between vector-like quarks and SM quarks needed to realize such an effective coupling in the first place. The mixing between the vector-like quarks and SM quarks is parametrised in our model by the ratio  $M^{\Psi d}_{ij}/M^\Psi_i$. As we will discuss below, given the lower limit on leptoquark mass from direct collider searches and the current limits from $B_s-\bar{B}_s$ mixing, we will need $M_{\psi d}/M_{\psi} \sim  0.1$ to explain the anomalous $B\rightarrow K+ \slashed{E}$ data, for a perturbative Yukawa coupling $Y$.  Such a value for $M_{\psi d}/M_{\psi}$ is not extremely hierarchical, given in our model, both the scales are proportional to the PQ scale.

\begin{figure}[t!]
\centering
\includegraphics[width=0.5 \textwidth]{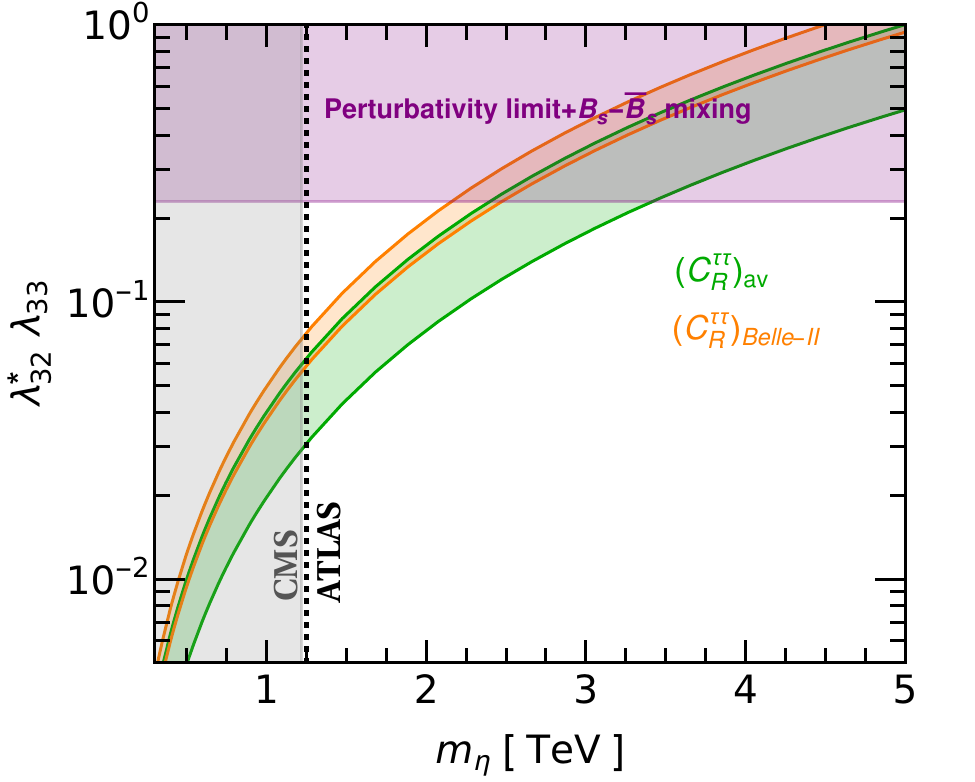} \hspace{-0.6cm}
\includegraphics[width=0.5 \textwidth]{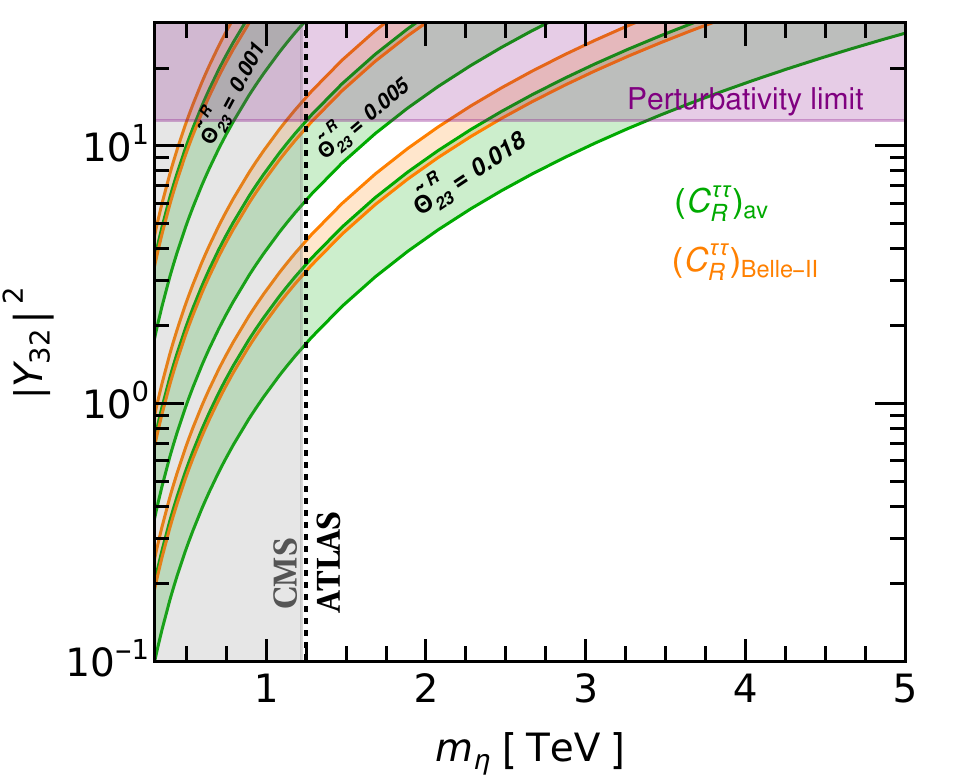} 
		\caption{\footnotesize Allowed regions in  $\lambda^*_{32} \lambda_{33}$  ($|Y_{32}|^2$) vs $m_{\eta}$ in the left (right) panel. The green and orange regions are consistent with the average value of $R^\nu_{K^+}$ and the latest measurement from Belle II. The exclusion regions using a vertical gray band and black dotted line arise from CMS~\cite{CMS:2023qdw} and ATLAS~\cite{ATLAS:2021jyv}, respectively (see text for details).}
		\label{fig:LQ} 
	\end{figure}

Given the assumed flavour structure of the effective coupling in Eq.~\eqref{eq:mfex6}, $B^0_s-\overline{B^0_s}$ meson mixing mediated via the flavour violating coupling of Higgs (see Appendix \ref{app:FCNCs}) provide constraints on the combination of mixing parameters~\cite{Alonso-Alvarez:2023wig}
\begin{equation}\label{eq:MesonMixing}
\tilde{\Theta}^R_{23}\equiv\sum_{i=1,2}{\Theta_R}_{i2}{\Theta_R^{\dagger}}_{i3} =\sum_{i=1,2}\frac {M^{\Psi d}_{i2} M^{\Psi d}_{i3} } {M^{\Psi^2}_{i}} \lesssim 0.018 \,,
\end{equation}
where $i$ labels the $\Psi$ species. There is also a constraint on the combination $\left({\lambda}^{\ast}_{i3}\right)^{ 2}{\left(\lambda_{i2}\right)^2}$, contributing to $B^0_s-\overline{B^0_s}$ via box-diagrams mediated by the scalar leptoquark and SM leptons (charged leptons and neutrinos). However, such contributions are lepton-mass suppressed (in addition to the mixing suppression due to vector-like quark and SM quark mixing) and are not expected to provide any useful constraints, see e.g.~\cite{Becirevic:2015asa} for a discussion on the relevant constraints. 
In the left panel of Fig.~\ref{fig:LQ}, we show allowed regions in the relevant scalar leptoquark coupling parameter versus its mass, $m_{\eta}$. The green and orange bands represent the regions preferred by the global average and only the latest measurement by Belle-II for $B\to  K E_{\text{miss}}$, c.f. Fig.~\ref{fig:CLCRplot}. The horizontal shaded region is excluded by the constraints from the neutral $B_s-\bar{B}_s$ mixing when we impose the perturbativity limit on the Yukawa couplings: $\vert Y_{32}\vert,\vert Y_{31}\vert \leq \sqrt{4\pi}$.

Leptoquarks may also be directly produced at the LHC, e.g. through $gg\,(q\bar q)\to \eta^\dagger\eta$. Both the ATLAS and CMS collaborations have searched for this process in various channels involving different quark/lepton final-states, $\mathrm{\eta}^\dagger\mathrm{\eta}\to q\bar q \ell\bar \ell,\,q\bar q\nu\bar \nu$. These searches lead to model-independent bounds on the mass and branching fractions of the leptoquarks. Note that, for couplings $\lambda_{32}$ and $\lambda_{33}$ less than order unity, the pair production is dominated by QCD interactions. It is important to note that, in our model, the effective coupling of the scalar leptoquark $\eta$ to SM quarks involves the mixing of vector-like and SM quarks. Therefore, unlike conventional leptoquark phenomenological literature, wherever a vertex involving coupling of $\eta$ with SM-quark and -lepton is involved, there is an additional mixing suppression factor. Assuming a 100\% branching of leptoquark into third-generation SM quarks and leptons, the current results from ATLAS~\cite{ATLAS:2021jyv} exclude masses up to approximately 1.25 TeV, while CMS~\cite{CMS:2023qdw} results exclude masses up to around 1.22 TeV, at the 95\%  confidence level. These limits are indicated by the vertical black dotted line and grey band, respectively, in Fig.~\ref{fig:LQ}, left panel. Note that single leptoquark exchange in t-channel and non-resonant production of leptoquarks are subdominant to pair production in the region of interest~\cite{CMS:2023qdw}.

In the right panel of Fig.~\ref{fig:LQ}, we show the relevant region in the scalar leptoquark Yukawa coupling $\vert Y_{32}\vert^2$ versus $m_{\eta}$. The colour coding is the same as in the left plot. Here, three distinct green (and orange) bands represent the allowed regions resulting from different benchmark values of the vector-like and SM quark mixing parameter $\tilde{\Theta}^R_{23}$. We note that the perturbativity requirement, together with the LHC direct search limits, already exclude the benchmark case where the mixing $\tilde{\Theta}^R_{23}\simeq 0.001$, capable of explaining the Belle-II excess. However, for larger value of the mixing $\tilde{\Theta}^R_{23}$ still allowed by neutral $B_s-\bar{B}_s$ mixing, the possibility to explain the excess $B\to  K + E_{\text{miss}}$ events at Belle-II remains viable for $1.25\; \text{TeV}<m_{\eta}< 3.36\; \text{TeV}$. The future measurements for neutral $B_s-\bar{B}_s$ mixing will, therefore, play a crucial role in probing our scenario. Another very interesting point to note is that the operators in Eqs. (\ref{LEFTvector}) and (\ref{eq:Lbsnunu-model}) can also lead to neutral mode transition $B^0 -> K^0*$ via an axial-vector term. The measurement of this mode would further impact the potential parameter space consistent with $B^+\to K^+ + E_{\text{miss}}$ data. A detailed study of the potential impact of such a potential measurement is beyond the scope of the current article. However, in Ref. \cite{Becirevic:2023aov}, some detailed discussion can be found along this direction. 

Moreover, the coupling combination ${\lambda}_{3 2}$ and ${\lambda}_{3 3}$ can give rise to new contributions to $B_s\to  \tau \tau$, which is predicted in the SM to have the branching fraction $\text{Br}(B_s\to  \tau \tau)=7.7(5) \times 10^{-7}$~\cite{PDG}. This mode is currently being searched by the LHCb and Belle-II collaborations~\cite{Belle-II:2018jsg} and can provide useful constraints on our scenario in the future. Another interesting mode of interest is the  $b \rightarrow s \tau \tau$ semi-leptonic meson decays, which is currently being attempted to be measured in experiments like LHCB and Belle-II)~\cite{Belle-II:2018jsg}. Once the branching fractions are available from the measurements, it will be interesting to explore this model further to check the consistency of the current anomaly with $b \rightarrow s \tau \tau$ data. The leptonic decay of $\Upsilon(nS)\to  \tau \tau$ is also relevant for the ${\lambda}_{3 3}$  coupling; however, such a mode is not a very clean probe because of large tree-level electromagnetic contributions~\cite{Becirevic:2016oho}.
\subsection{Comments on other phenomenologically interesting flavour observables}
Even though not required to explain the enhanced $B^+\to K^+ + E_{\text{miss}}$ rate at Belle-II, if the effective couplings ${\lambda}$ to muons (i.e. ${\lambda}_{22}$ and ${\lambda}_{23}$) are also non-vanishing, our model can give rise to several interesting low-energy observables involving the second and third generation of quarks and leptons. 
Indeed, note that if at least one of the couplings is non-zero in the second and third rows of  Eq.~\eqref{eq:mfex4}, then two non-vanishing light neutrino masses are generated at the two-loop level as we will discuss in the next section. Many low-energy observables involving the second and third generation of quarks and leptons observables are currently being searched for in the current generation of experiments that may potentially improve the sensitivities in the near future. 

For example, there can be charged lepton flavour violating decays, such as $\tau\to  \mu \gamma$. This proceeds at the one-loop level mediated by the effective scalar leptoquark $\eta$ coupling with SM quarks propagating in the loop. The vector-like quarks $\Psi$ can also directly propagate in the loop together with $\eta$. The importance of this contribution is dictated by the heaviness of the vector-like quarks $\Psi$. Since the mass of $\Psi$ is proportional to the PQ breaking scale, it can range anywhere between the PQ breaking scale (for order unity couplings) down to several TeV, for small couplings $\mathcal{O}(10^{-9})$. The mass of $\Psi$ is not fixed by the rate for $b\to s \nu_{\tau}\overline{\nu_{\tau}}$, as it is only sensitive to the ratio $\frac{M^{\Psi d}}{M^{\Psi}}$. 
If the mass of $\Psi$ lies in the TeV scale then the contribution from the $\Psi$-mediated diagrams becomes particularly important. On the other hand, for large $\Psi$ masses close to the PQ-breaking scale, we can safely ignore contributions from diagrams involving the vector-like quarks and only consider those involving the SM quarks instead, leading to the $\tau \to \mu \gamma$ branching fraction~\cite{Becirevic:2016oho}

\begin{equation}
\text{Br}(\tau\to \mu \gamma)=\frac{\alpha}{(m_\tau^2-m_\mu^2)^3}{4 m_\tau^3 \Gamma_\tau} \left(  \lambda'_{23} {\lambda^{'\ast}_{33}} \frac{3 m_b^2 m_\tau}{96 \pi^2 M_\eta^4} \left[\frac{5}{2}-\log\left(\frac{M_\eta^2}{m_b^2}\right)\right]\right)^2,
\end{equation}
assuming that the bottom quark provides the dominant loop contribution. Here $m_\ell$ with $\ell=\mu, \tau$ denotes the charged lepton masses, $\Gamma_\tau$ is the $\tau$ decay width, $m_b$ denotes the bottom quark mass and $\alpha$ is the fine-structure constant. 
The current upper limit from Babar experiment on the branching fraction $\text{Br}(\tau\to  \mu \gamma)<4.4\times 10^{-8}$~\cite{BaBar:2009hkt} leads to a rather weak constraint.

If ${\lambda}_{2 2}$ and ${\lambda}_{3 2}$ are both present and sizable in our model, then $\tau\to  \mu \phi$ can provide useful constraints. The relevant branching fraction is given by~\cite{Becirevic:2016oho}
\begin{equation}
\text{Br}(\tau\to\mu \phi)=\frac{f_\phi^2 m_\phi^4}{256 \pi m_\tau^3 \Gamma_\tau}\left|\frac{\lambda'_{32}{\lambda^{' \ast}_{22}}}{m_\eta^2} \right|^2 \left[-1+\frac{(m_\mu^2+m_\tau^2)}{2 m_\phi^2}+\frac{(m_\mu^2-m_\tau^2)^2}{2 m_\phi^4}\right]\lambda^{1/2}(m_\phi^2,m_\tau^2,m_\mu^2),
\end{equation}
where $f_\phi$ and $m_\phi$ are the decay constant and mass of the $\phi$-meson. The current experimental upper limit $\text{Br}(\tau\to  \mu \phi)<8.4 \times 10^{-8}$ leads to a constraint
\begin{equation}
\sum_{i=1,2}\frac {\lambda'_{22} {\lambda^{' \ast}_{32}} } {\left(\frac{M_\eta}{\text{TeV}}\right)^2} \lesssim 0.036
\end{equation}
Note that, in the mass basis after diagonalising the SM charged current interactions $\lambda$ and $\lambda'$ couplings are related by the unitary lepton rotation matrix, c.f. Eq.~\eqref{eq:coupmb1}

If ${\lambda}_{2 2}$ and ${\lambda}_{2 3}$ are both present and sizable in our model, then the current measurements of the observables $B_s\to  \mu\mu$, $B\to  K\mu^+\mu^-$, $B_s\to  \phi\mu^+\mu^-$ provides useful constraint, as these are in good agreement with the flavour universal SM predictions, up to some unresolved tensions in a few angular observables and total branching ratios~\cite{Gubernari:2022hxn,Ciuchini:2022wbq,Alguero:2023jeh,Wen:2023pfq}. However, we stress that these couplings are not required for explaining the enhanced $B^+\to K^+ + E_{\text{miss}}$ rate at the Belle-II experiment. 

The current world average of the universality ratios for the charged current $B$-decay modes $R_D^{(*)}\equiv\text{Br}(B\rightarrow D^{(*)}\tau \nu)/\text{Br}(B\rightarrow D^{(*)}\ell \nu)$ also show deviation from the SM predictions~\cite{HFLAV:2022esi}. However, the scalar leptoquark $\eta$ is not suitable for explaining this deviation, see e.g.~\cite{Becirevic:2024pni}. On the other hand, the scalar $\chi$ might do the job~\cite{Becirevic:2024pni,Carvunis:2021dss,Shi:2019gxi,Hati:2015awg,Fajfer:2012jt} provided additional vector-like quark doublet(s) and vector-like singlet up-type quark(s) are introduced in the model facilitating couplings of $\chi$ with SM up- and down-type quarks via mass-mixing. However, a detailed exploration along this line is beyond the scope of the current work.

\section{Two-loop neutrino mass generation}
\label{sec:numass}

\begin{figure}[t!]
		\centering
         \includegraphics[scale=0.43]{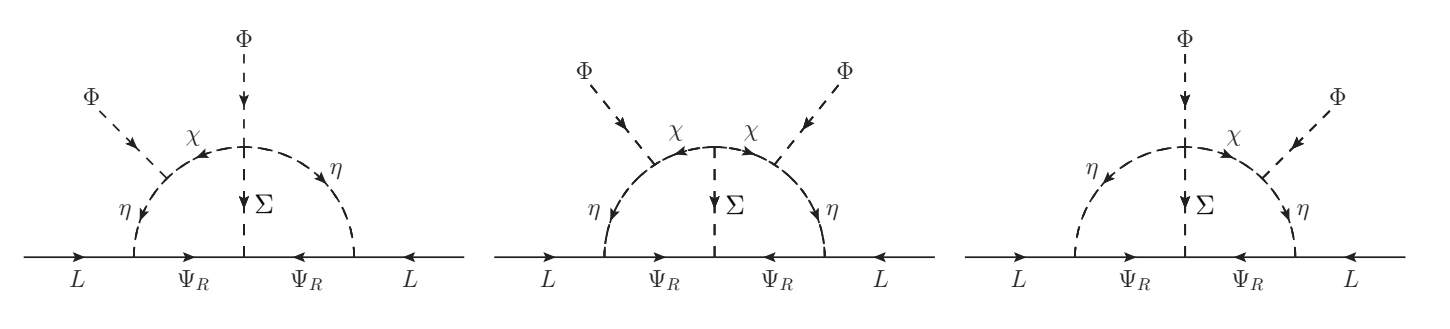}
		\caption{Two-loop neutrino masses in the symmetry basis.}
		\label{fig:nudiags}
	\end{figure}

One of the most attractive features of our scenario for explaining the Belle-II anomalous measurement is that it implements an axion solution to the strong CP problem with colour-mediated neutrino masses.
 Neutrino mass generation in our framework is depicted by the two-loop diagrams in Fig.~\ref{fig:nudiags}. 
Notice that the centre of each neutrino mass diagram involves the interaction term $\frac{Y^{\Sigma}_{ab}}{2}  {\Psi^T_{aR}}\,C\,\Sigma\Psi_{bR}$.
If $\Sigma$ was also a triplet of $SU(3)_c$, the corresponding Yukawa coupling matrix would be anti-symmetric in flavour space, leading to vanishing neutrino masses.
However, in our proposed model, $\Sigma$ transforms as an anti-sextet under $SU(3)_c$, so that the Yukawa matrix is symmetric in flavour space, allowing for non-zero neutrino masses, as we see next. Yet, in the minimal case under consideration with $n_\Psi = 2$, one neutrino remains massless as a consequence of the missing partner nature of our mass generation mechanism in this case, see discussion in~\cite{Schechter:1980gr}.

In order to calculate the two-loop masses, it is convenient to move to the mass basis. After electroweak symmetry breaking, the coloured scalars with electric charge equals $-1/3$,  in the basis $( \chi,\, \eta^{-1/3})$, mix according to:
\begin{eqnarray}
    \bm{M}^2 &=& \begin{pmatrix}
        m^2_{\chi} & \frac{\kappa v}{\sqrt{2}} \\
        \frac{\kappa v}{\sqrt{2}} & m^2_{\eta} 
    \end{pmatrix}\,,\\m^2_{\chi} &=&\frac{1}{4} (\lambda_{\Phi \chi} v^2 + \lambda_{\chi \sigma} v_\sigma^2)+ \mu_{\chi}^2\,,\nonumber\\
    m^2_{\eta} &=&\frac{1}{4} \left[(\lambda_{\Phi \eta}+\tilde{\lambda}_{\Phi \eta}) v^2 + \lambda_{\eta \sigma} v_\sigma^2\right]+ \mu_{\eta}^2\,.\nonumber
\end{eqnarray}
The mass states, $\phi_{1,2}$, are then related to the symmetry states via
\begin{eqnarray}
      \begin{pmatrix}
        \phi_1 \\ \phi_2
    \end{pmatrix}= \begin{pmatrix}
        \cos \theta_{LQ} & -\sin \theta_{LQ} \\
        \sin \theta_{LQ} & \cos \theta_{LQ}
    \end{pmatrix}\begin{pmatrix}
        \eta^{-1/3} \\ \chi
    \end{pmatrix}\, \quad \mbox{with} \quad 
    \tan{2 \theta_{LQ}} = \frac{\sqrt{2}\,\kappa\, v}{ m^2_\eta - m^2_{\chi} }\,.
\end{eqnarray}
If one assumes, for instance, that $|\kappa|$ is no larger than $v= 246$ GeV, while $m_{\eta}$ and $m_{\chi}$ are at least TeV-scale masses, the mixing angle is small, $\theta_{LQ}\ll 1$, and the physical masses are of approximately given by $m_{\phi_1} \simeq m_\chi$ and $m_{\phi_2} \simeq m_\eta$. As for the other scalar in the loop, $\Sigma$, with electric charge of $2/3$, it remains unmixed at tree level, with
\begin{eqnarray}\label{eq:SigmaMass}
m_{\Sigma}^2 =\frac{1}{4} (\lambda_{\Phi \Sigma} v^2 + \lambda_{\sigma\Sigma} v_\sigma^2)+ \mu_{\Sigma}^2\,.
\end{eqnarray}

Similarly, we can write the quarks in the loop in terms of the physical fields using Eq. (\ref{eq:qtrans})
\begin{eqnarray}
    \bm{\Psi}_{R} \simeq (\bm{\mathcal{I}}_2-\bm{\Theta}_R^\dagger \bm{\Theta}_R)^{1/2}\,\bm{V}^{\Psi}_R \,\bm{\Psi^\prime}_{R} - \bm{\Theta}_R^\dagger\, \bm{V}^d_R\, \bm{d^\prime}_R\,.
\end{eqnarray}
 
The two-loop neutrino masses arising from the left (L), centre (C) and right (R) diagrams in Fig. \ref{fig:nudiags} can then be written as 
\begin{eqnarray}\label{eq:numasses}
    \mathcal{M}^{\nu\,(L)}_{{ij}} &=&  6\,\frac{\lambda\,v}{\sqrt{2}}\,
    s_{\theta_{LQ}} c_{\theta_{LQ}} \sum_{\alpha, \beta, N} \tilde{Y}_{i\alpha}^*\,\tilde{Y}^\Sigma_{\alpha \beta}\,(\tilde{Y}_{j \beta})^\dagger\,\left( -c_{\theta_{LQ}}\, \mathcal{I}^{N}_{\alpha\beta,11}-s_{\theta_{LQ}}\,\mathcal{I}^{N}_{\alpha\beta,12}+c_{\theta_{LQ}}\,\mathcal{I}^{N}_{\alpha\beta,21}+ s_{\theta_{LQ}}\,\mathcal{I}^{N}_{\alpha\beta,22}\right)\,,\nonumber\\
    \mathcal{M}^{\nu\,(C)}_{ij} &=& 3\,\mu\,
    (s_{\theta_{LQ}} c_{\theta_{LQ}})^2 \,\sum_{\alpha, \beta, N}\tilde{Y}_{i\alpha}^*\,\tilde{Y}^\Sigma_{\alpha \beta}\,(\tilde{Y}_{j\beta})^\dagger\,\left(  \mathcal{I}^{N}_{\alpha\beta,11}-\mathcal{I}^{N}_{\alpha\beta,12}-\mathcal{I}^{N}_{\alpha\beta,21}+ \mathcal{I}^{N}_{\alpha\beta,22}\right)\,,\\
    \mathcal{M}^{\nu\,(R)}_{ij} &=& 6\,\frac{\lambda\,v}{\sqrt{2}}\,
    s_{\theta_{LQ}} c_{\theta_{LQ}} \sum_{\alpha, \beta, N} \tilde{Y}_{i\alpha}^*\,\tilde{Y}^\Sigma_{\alpha \beta}\,(\tilde{Y}_{j \beta})^\dagger\,\left( -c_{\theta_{LQ}}\, \mathcal{I}^{N}_{\alpha\beta,11}-s_{\theta_{LQ}}\,\mathcal{I}^{N}_{\alpha\beta,21}+c_{\theta_{LQ}}\,\mathcal{I}^{N}_{\alpha\beta,12}+ s_{\theta_{LQ}}\,\mathcal{I}^{N}_{\alpha\beta,22}\right)\,,\nonumber
\end{eqnarray}
where the numerical prefactors above take into account the possible independent colour contractions and symmetry factors and
\begin{eqnarray}\label{eq:Iints}
    \mathcal{I}_{\alpha\beta,ab}^{N} &=& \int \frac{d^4 k}{(2\pi)^4}\,\int \frac{d^4 p}{(2\pi)^4} \frac{N}{(k^2-m_{\phi_a}^2) (p^2-m_{\phi_b}^2)[(p+k)^2-m_{\Sigma}^2] (k^2-m_{B^\prime_\alpha}^2) (p^2-m_{B^\prime_\beta}^2) }\,,
\end{eqnarray}
with the indices $\alpha,\beta = 1,\cdots,5$ running on the quark mass basis $\bm{B^\prime} = (\bm{d^\prime}\,\bm{\Psi^\prime})^T$, as defined in Eq. (\ref{eq:qtrans}),  $a,b=1,2$ and the numerator $N = \{2 m_{B^\prime_\alpha} m_{B^\prime_\beta},p^2, k^2,-(p+k)^2\}$, with
\begin{eqnarray}
    \bm{\tilde{Y}}_{[3\times 5]} = \begin{pmatrix}
         - \bm{Y} \bm{\Theta}_R^\dagger & \bm{Y}
               \end{pmatrix}\quad \quad\quad\mbox{and}\quad\quad\quad
        \bm{\tilde{Y}^\Sigma}_{[5\times 5]} = \begin{pmatrix}
          \bm{\Theta}_R^* \bm{Y^\Sigma} \bm{\Theta}_R^\dagger & \bm{\Theta}_R^* \bm{Y^\Sigma}\\
           \bm{Y^\Sigma} \bm{\Theta}_R^\dagger & \bm{Y^\Sigma}
               \end{pmatrix}\,.
\end{eqnarray}
The solutions to the loop integrals are presented in Appendix \ref{app:LoopInts}. 

For the sake of illustration, we choose $\lambda \to 0$ in such a way that the central diagram in Fig. \ref{fig:nudiags} gives the dominant contribution to neutrino masses. We also assume the scenario where the lepton mixing matrix is associated with the charged lepton sector, as discussed in Sec. \ref{sec:modelex}; therefore, the neutrino mass matrix in Eq. (\ref{eq:numasses}) must be diagonal. Moreover, since the heavy quarks, $\Psi_{a}^\prime$, get masses proportional to the PQ scale, while the scalar mediators are assumed to be around the TeV scale, the diagrams mediated by $\Psi_{a}^\prime$ are highly suppressed when compared to those mediated by the SM quarks.
In this case 
$\bm{\mathcal{M}}^{\nu}$ can be written as
\begin{eqnarray}
    \bm{\mathcal{M}}^{\nu} &=& \mbox{diag}(0, m_2^\nu, m_3^\nu) \simeq  \bm{Y}^* \bm{\Theta}_R^T\,\left(\bm{\Theta}_R^* \bm{Y^\Sigma} \bm{\Theta}_R^\dagger \star \bm{\tilde{\mathcal{I}}} \right)\,\bm{\Theta}_R \bm{Y}^\dagger\,,\\
    \mbox{with} \quad \quad 
    \tilde{\mathcal{I}}_{\alpha\beta} &=& 3\,\mu\,
    (s_{\theta_{LQ}} c_{\theta_{LQ}})^2 \sum_N \left(  \mathcal{I}^{N}_{\alpha\beta,11}-\mathcal{I}^{N}_{\alpha\beta,12}-\mathcal{I}^{N}_{\alpha\beta,21}+ \mathcal{I}^{N}_{\alpha\beta,22}\right)\,,\nonumber
\end{eqnarray}
and $\star$ represents the Hadamard, or element-wise, matrix product. Then, after multiplying both sides by $(\bm{\mathcal{M}^{\nu\,}})^{-1/2} = \mbox{diag}( 0, \sqrt{1/m_2^\nu}, \sqrt{1/m_3^\nu} )$, we obtain
\begin{eqnarray}\label{eq:RotationMatrix} \small
    \mbox{diag}(0, 1,1 ) =  \left[ (\bm{\mathcal{M}^{\nu\,}})^{-1/2}\bm{Y}^* \bm{\Theta}_R^T\, \left(\bm{\Theta}_R^* \bm{Y^\Sigma} \bm{\Theta}_R^\dagger \star \bm{\tilde{\mathcal{I}}} \right)^{1/2}\right]\,\left[(\bm{\mathcal{M}^{\nu\,}})^{-1/2}\bm{Y}^* \bm{\Theta}_R^T\, \left(\bm{\Theta}_R^* \bm{Y^\Sigma} \bm{\Theta}_R^\dagger \star \bm{\tilde{\mathcal{I}}} \right)^{1/2}\right]^T\,.
\end{eqnarray}
A solution to this equation is given by
\begin{eqnarray}\label{eq:NuMassEq}
    \bm{R} \equiv  \begin{pmatrix}
        0 & 0  \\   
        \cos \theta_R & \sin \theta_R  \\
        -\sin \theta_R & \cos \theta_R
    \end{pmatrix}\, = \left[ (\bm{\mathcal{M}^{\nu\,}})^{-1/2}\bm{Y}^* \bm{\Theta}_R^T\, \left(\bm{\Theta}_R^* \bm{Y^\Sigma} \bm{\Theta}_R^\dagger \star \bm{\tilde{\mathcal{I}}} \right)^{1/2}\right]\,,
\end{eqnarray}
with $\theta_R$ being a generic complex angle. Finally, Eq.~\eqref{eq:RotationMatrix} can be solved for the elements of $\bm{Y}^*$ in terms of the neutrino masses and the other parameters as follows
\begin{eqnarray}\label{eq:Yparametrisation}
    \mbox{diag}(0, 1,1 )\, \bm{Y}^* = \begin{pmatrix}
        0 & 0 \\
        Y_{21}^* & Y_{22}^* \\
        Y_{31}^* & Y_{32}^* \\
    \end{pmatrix} = {\bm{\mathcal{M}^{\nu}}}^{1/2}\, \bm{R} \, \left\{\bm{\Theta}^T_R \left[\left(\bm{\Theta}^*_R \bm{Y}^\Sigma \bm{\Theta}^\dagger_R\right) \star \,
 \bm{\tilde{\mathcal{I}}}\right] \bm{\Theta}_R \right\}^{-1/2}\,.
\end{eqnarray}

\begin{figure}[t!]
		\centering
		\includegraphics[width=0.8 \textwidth]{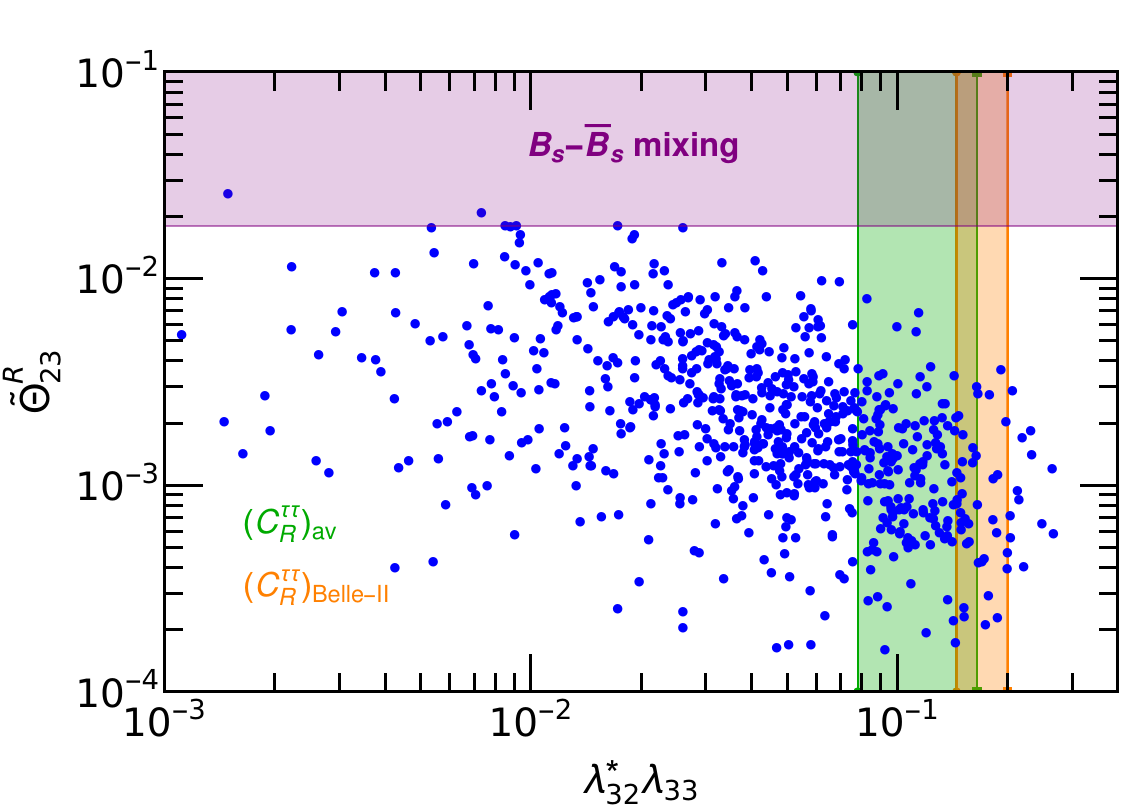} 
		\caption{\footnotesize  The blue points reproduce neutrino oscillation mass-squared requirements~\cite{deSalas:2020pgw} for  a benchmark value $m_\eta = 2$ TeV.  The green and orange bands represent the parameter regions that explain the combined and Belle-II only $B^+ \to K^+ + E_{\mbox{miss}}$ anomalous measurement, respectively.}  
		\label{fig:mnuAnomaly} 
	\end{figure}
%
For illustration, we now perform a concrete numerical analysis in terms of model parameters that could explain neutrino oscillations using Eq. \eqref{eq:Yparametrisation}.
In Fig.~\ref{fig:mnuAnomaly}, we present our results using blue points in the $\tilde{\Theta}^R_{23}$ vs $\lambda_{32}^*\, \lambda_{33}$ plane. These points are consistent with the neutrino mass-squared differences reported by the global analysis of neutrino oscillation data~\cite{deSalas:2020pgw}.
For the scan, we assumed $m_\eta,m_\Sigma \simeq 2$ TeV and $m_\chi \simeq  4$ TeV with a small mixing angle $\theta_{LQ} \lesssim 4 \times 10^{-3}$.  
As for the neutrino mass mediators, the SM quarks give the main contribution, since the exotic quark masses are around the PQ scale.  The horizontal magenta band is excluded by $B^0_s-\overline{B^0_s}$ meson mixing constraints, whereas the vertical green (orange) bands represent the $\lambda_{32}^* \lambda_{33}$ regions in which the average (Belle-II only) $B^+ \to K^+ + E_{\mbox{miss}}$ anomalous measurement is explained via $\eta$ exchange, following the discussion in Sec. \ref{sec:modelex}.

\section{Flavor violating QCD axion}\label{sec:AxionPheno}
In this section, we summarise some of the salient features of axion phenomenology in our model~\cite{Batra:2023erw}, including the flavour-violating (FV) interactions of axions. By construction, the SM fermions are not charged under the PQ-symmetry~\cite{Kim:1979if,Shifman:1979if}, leading to no direct coupling with the axion. Nevertheless, the mixing between the vector-like quarks $\Psi_{a}$, charged under PQ, and the SM down-type quarks, shown in Eq. (\ref{eq:qtrans}), induces couplings between the latter and the axion. Thus, the FV interactions between the light down-type quarks and the axion can be written as 

\begin{eqnarray}\label{eq:LaFV}
    \mathcal{L}_a^{\rm FV} &=& \frac{ \partial_\mu a}{f_a}\, \overline{d^\prime_{i}}\,\gamma^\mu \,\left( c^V_{ij}  + c^A_{ij} \gamma_5 \right)\, d^\prime_{jR}
    \,, \\
    \mbox{with}\quad\quad && c^V_{ij} = c^A_{ij} = c_{ij} =\frac{1}{2}\mathcal{Q}_R\tilde{\Theta}^R_{ij}\,.\nonumber
\end{eqnarray}
where $\tilde{\Theta}^R_{ij}$ is defined in Appendix \ref{app:FCNCs} and $\mathcal{Q}_R$ is the PQ charge of $\Psi_{aR}$ as given in Table~\ref{tab:general}. Here $f_a$  is the axion decay constant and $c^{V,A}_{\psi_i \psi_j}$ represent the vector and axial couplings, which are Hermitian matrices in flavour space.

In our model, $a$ is the QCD axion, so that its mass is inversely proportional to $f_a$~\cite{GrillidiCortona:2015jxo}
\begin{equation}\label{eq:FaMa}
    m_a = 5.691\times10^{-6}\,\mathrm{eV} \, \left( \frac{10^{12}\,\mathrm{GeV}}{f_a} \right).
\end{equation}
 As discussed above, the axions can couple only with the right-handed down-type quarks via the mixing of vector-like and SM quarks. Moreover, in the minimal scenario to account for an enhanced $B^+\to K^+ + E_{\text{miss}}$ rate at Belle-II, 
the relevant couplings in Eq.~\eqref{eq:LaFV} can be written as
\begin{equation}\label{eq:CvCa}
    c_{bs}^V = c_{bs}^A = \frac{1}{2} \mathcal{Q}_R \,\tilde{\Theta}^R_{23} \,.
\end{equation}
Following Eq.~\eqref{eq:MesonMixing}, we express the mixing parameter for our benchmark case as
\begin{equation}
\tilde{\Theta}^R_{23} =  \frac {M^{\Psi d}_{22} M^{\Psi d}_{23} } {M^{\Psi^2}_{2}}   =  \frac {M^{\Psi d}_{22} M^{\Psi d}_{23} } { (Y_{22}^{\Psi}f_a)^2} \,, 
\end{equation}
where the relation between $f_a$ and $m_a$ is given by Eq.~\eqref{eq:FaMa}. Recall that the constraints from the neutral $B_s-\bar{B}_s$ mixing lead to $\tilde{\Theta}^R_{23}<0.018$. Note that the mass term $M^\Psi_2$ appears after spontaneous breaking of $U(1)_{PQ}$ and electroweak symmetry, as given in Eq.~\eqref{eq:mixmat}, so we take $M_2^{\Psi} = Y_{22}^{\Psi} \, f_a$ with $Y_{22}^{\Psi}$ of $\mathcal{O} (1)$.
\begin{figure}[t!]
\centering
\includegraphics[width=0.75 \textwidth]{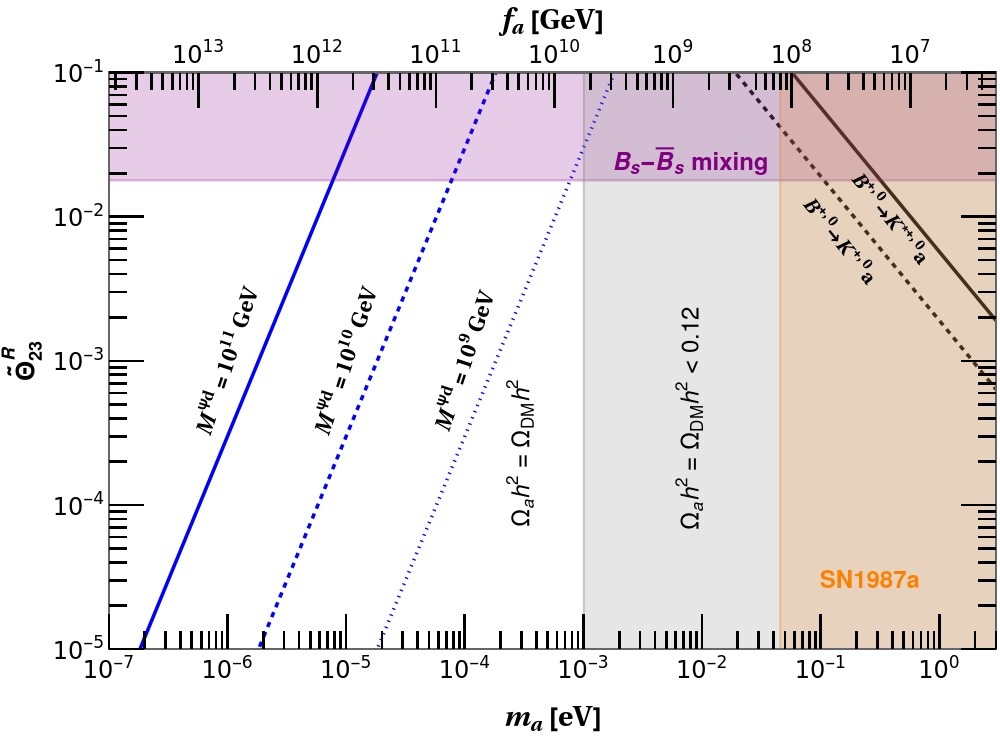} 
		\caption{\footnotesize The limits are displayed in the plane of $m_a$ (bottom axis) and $f_a$ (top axis) versus the absolute value of $\tilde{\Theta}^R_{23} $. The allowed region, where the QCD axion can explain the observed dark matter relic abundance, is depicted in white. The blue lines are simulated based on three benchmark values of model parameters. Different exclusion regions are presented using the purple, grey, and light-orange bands for $B^0_s-\overline{B^0_s}$ meson mixing~\cite{Alonso-Alvarez:2023wig}, DM under-abundance, and SN1987a~\cite{Carenza:2019pxu}, respectively (see text for details). 
  }\label{fig:Axion} 
  \end{figure}
Fig.~\ref{fig:Axion} gives the allowed region in terms of the axion mass $m_a$ (or the corresponding decay constant $f_a$) and the absolute value of the mixing parameter $\tilde{\Theta}^R_{23}$. Here we took three mass parameter benchmarks, $M^{\Psi d}_{22} = M^{\Psi d}_{23} = M^{\Psi d}$, setting them to  $ 10^{11}$ GeV, $10^{10}$ GeV, and $10^{9}$ GeV. These cases are represented by the solid, dashed, and dotted blue lines, respectively.

One of the attractive features of the QCD axion is that it can explain the relic Dark matter abundance of the Universe. 
In the white region in Fig.~\ref{fig:Axion}, the QCD axion can constitute the observed dark matter relic density~\cite{Adams:2022pbo, Buschmann:2021sdq, Gorghetto:2020qws, Klaer:2017ond, Kawasaki:2014sqa}, whereas in the grey-shaded region, the QCD axion leads to an under-abundant axion dark matter~\footnote{See~\cite{Ghosh:2022rta,Ghosh:2023xhs,Ghosh:2024boo} for some recent studies on KSVZ type models with potentially interesting collider phenomenology.}.
Furthermore, the white region satisfying $\tilde{\Theta}^R_{23}<0.018$ and meeting all other constraints remains consistent with explaining the Belle-2 anomaly.

In contrast, the light-orange vertical region denotes the constraints on the axion-nucleon couplings derived from SN1987a~\cite{Carenza:2019pxu}. The model-independent axion-nucleon couplings, induced by the axion-gluon interactions, are $C_{an} = -0.02$ and $C_{ap} = -0.47$ \cite{Carenza:2019pxu,DiLuzio:2020wdo}. The constraint from Supernova is dominantly $0.29 \times g_{ap}^2< 3.25 \times 10^{-18}$, where $g_{ap} \sim m_p C_{ap}/f_a$ with $m_p$ being the mass of proton. This translates into a lower bound in the ballpark of $f_a \sim 10^8$ GeV. Other than the constraints from the $B^0_s-\overline{B^0_s}$ meson mixing as we already discussed, the BaBar collaboration~\cite{BaBar:2013npw} conducted searches for $B^{+,0}\to K^{+,0} \, + a$ and $B^{+,0}\to K^{* +,0} \, +a$. These searches also provide constraints on the flavor-violating coupling $\tilde{\Theta}^R_{23}$, see e.g.~\cite{MartinCamalich:2020dfe}, which we show as the dashed black lines in Fig.~\ref{fig:Axion}.
To obtain these lines,  we find the values of $ c_{bs}^{V,A} (\equiv f_a/ F_{bs}^{V,A} ) $ from Tab. 3 of  ~\cite{MartinCamalich:2020dfe} and then utilizing Eq. \ref{eq:FaMa} and \ref{eq:CvCa}, the results are expressed in the adopted parameter space.

Next-to-leading-order chiral Lagrangian techniques can be used to obtain a relation between axion-photon coupling ($g_{a \gamma \gamma}$) and PQ breaking scale $f_a$ and is given by~\cite{GrillidiCortona:2015jxo}
\begin{align}
    g_{a \gamma \gamma} &= \frac{\alpha_e}{2 \pi f_a} \left[\frac{E}{N} - 1.92(4) \right] \; ,
    \label{eq:gagg}
\end{align}
where $E$ ($N$) represents the model-dependent electromagnetic (colour) anomaly factor. Thus, in our scenario (see Table~\ref{tab:general}) the ratio $E/N$ for $\Psi \sim \, (3, 1, -1/3)$ turned out to be $2/3$. The prediction of such a scenario mimics the outcome of a DFSZ II-type setup~\cite{Batra:2023erw}. The haloscope experiments, such as upgraded ADMX~\cite{Stern:2016bbw} and upcoming  MADMAX~\cite{Beurthey:2020yuq}, could potentially probe the prediction of our scenario through a direct search.
\section{Conclusions and outlook}
\label{sec:ALP}


We studied the possibility of explaining the recent Belle-II excess of $B^+\to K^+ + E_{\text{miss}}$ events within a KSVZ-type axion model proposed in~\cite{Batra:2023erw}. The model implements a Peccei-Quinn (PQ) symmetry, leading to a QCD-axion capable of addressing the strong CP problem and providing the dark matter relic abundance. The model contains a scalar PQ-charged leptoquark, vector-like quarks and some additional coloured scalars, which can naturally lead to two-loop radiative light neutrino masses, Fig.~\ref{fig:nudiags}. 
The scalar leptoquark only interacts with SM leptons and vector-like quarks at the tree level. The mass-mixing of vector-like quarks with the SM quarks leads to effective coupling of the PQ-charged scalar leptoquark with the SM quarks providing a potential explanation of anomalous $B^+\to K^+ + E_{\text{miss}}$ events via a vector current $B^+\to K^+ \nu_\tau \nu_\tau$, Fig.~\ref{fig:B2Knunu}. We discussed the flavour structure of the scalar leptoquark couplings as required for the above, and also how such a structure leads to a correct light neutrino mass spectrum consistent with the neutrino oscillation data. These are shown in Figs.~\ref{fig:LQ} and~\ref{fig:mnuAnomaly}, respectively.
We also commented on possible probes of the axion couplings in our model. If the anomalous $B^+\to K^+ + E_{\text{miss}}$ events persist in future searches, then the model viability, in its minimal form, may be tested at a number of ongoing and future experiments involving the two heavier generations of charged leptons and quarks. Among them, improved neutral $B_s-\bar{B}_s$ oscillation limits will be crucial in probing the relevant parameter region where the model explains the Belle-II  $B^+\to K^+ + E_{\text{miss}}$ excess. 

We also note an interesting feature of our minimal construction: one of the three light neutrinos remains massless.
It emerges as reminiscent of the ``missing partner'' nature of the loop diagrams, first discussed in the context of the tree-level seesaw mechanism in Ref.~\cite{Schechter:1980gr}.
In such a scheme, no cancellation in the {$0\nu\beta\beta$ } decay amplitude is possible, even for the case of normal neutrino mass ordering~\cite{Reig:2018ztc,Barreiros:2018bju,Rojas:2018wym,Mandal:2021yph,Avila:2019hhv}.
This implies an absolute {$0\nu\beta\beta$ } decay lower bound and improved detection properties in the case of inverted mass ordering, see discussion in~\cite{Ding:2024ozt}.

Before closing, we mention a second possibility to explain the Belle-II  $B^+\to K^+ + E_{\text{miss}}$ anomaly by making the axion gain a soft mass term, breaking the PQ symmetry. In such an axion-like particle (ALP) scenario, the anomalous Belle-II measurement can be explained through the flavour violating $b \to s  a$ transition, where $a$ denotes the ALP. In Ref.~\cite{Altmannshofer:2023hkn, Bolton:2024egx}, it was reported that the minimization of the binned likelihood suggests a best-fit value of the ALP mass $m_a=2.1\pm 0.1$ GeV for the two-body decay $B\to  K^{(*)} a$. The remaining model ingredients remain the same, with the vector-like quarks playing a crucial role in inducing the flavour off-diagonal ALP couplings. The neutrino mass is generated in the same fashion as in our original model.

\begin{acknowledgements} 
This work was supported by Spanish Grants No. PID2023-147306NB-I00 and CEX2023-001292-S (MCIU/AEI/10.13039/501100011033), as well as Prometeo CIPROM/2021/054 (Generalitat Valenciana). C.~H. is supported by the CDEIGENT grant No. CIDEIG/2022/16 (funded by Generalitat Valenciana under Plan Gen-T).
\end{acknowledgements}

\begin{appendix}

\section{Tree-level flavour-violating quark interactions with $Z$, $h$ and $a$}\label{app:FCNCs}
As a consequence of the mixing between SM and heavy quarks, parametrised in terms of $\bm{\Theta}_L$ and $\bm{\Theta}_R$ in Eq. (\ref{eq:qmix}), flavour-changing neutral currents (FCNCs) arise in the down-type quark sector, mediated not only by the SM $Z$ boson and Higgs field $h$, but also by the axion $a$.

To identify the quark flavor-violating couplings of the $Z$ boson, we can start from the definition of the electroweak covariant derivative $D_\mu = \partial_\mu - i g T^a W^a_\mu - i g_Y Y B_\mu$, where $g$ and $g_Y$ are the $SU(2)_L$ and $U(1)_Y$ gauge couplings, respectively, $T^a$ and $Y$ the generators of such groups, and $W_\mu^a$ and $B_\mu$ the associated gauge fields. 
As usual, $W^3_\mu$ and $B_\mu$ can be rewritten in terms of the physical vector bosons $Z_\mu$ and $A_\mu$, by the standard combination given by the Weinberg angle $\theta_W$, and the combination $\bm{Q} = \bm{T^3} + \bm{Y}$ defining the electric charge operator. Thus, the contribution of $D_\mu$ proportional to the $Z$ boson can be written as
\begin{eqnarray}
    D_\mu =  \frac{-ig}{\cos{\theta_W}}\left(\bm{T^3} - \sin^2 \theta_W \bm{Q}\right)Z_\mu + \cdots
    \label{eq:ZCD}
\end{eqnarray}
Therefore, from Eqs. (\ref{eq:qtrans}), (\ref{eq:qmix}) and (\ref{eq:ZCD}), we find that the flavour-violating interactions mediated by the $Z$ boson are
\begin{eqnarray}\label{eq:FCNCZ}
    \mathcal{L}_{Z}^{\rm FV} = \frac{g}{2\cos{\theta_W}}\overline{d^\prime_{iL}}\,\gamma^\mu \,\tilde{\Theta}^L_{ij}\, d^\prime_{jL} Z_\mu\quad \mbox{with} \quad \bm{\tilde{\Theta}}^L = {\bm{V^d}_L}^\dagger \bm{\Theta}_L \bm{\Theta}_L^\dagger \bm{V}^d_L\,.
    \label{eq:ZFV}
\end{eqnarray}
In the scenario where the CKM matrix is ascribed to the up-type quarks, $\bm{V}^d_L = \bm{\mathcal{I}}_3$, we have that $\bm{\tilde{\Theta}}^L = \bm{\Theta}_L \bm{\Theta}_L^\dagger$.

Regarding the interactions with the Higgs boson $h$, the main contribution to flavour violation arises from the last Yukawa term in Eq. (\ref{eq:Yuk}). Upon using the expression in Eqs. (\ref{eq:qtrans}) and (\ref{eq:qmix}), and assuming that $\phi^0= (v+h)/\sqrt{2}$, we find 
\begin{eqnarray}\label{eq:FCNCh}
   \mathcal{L}_{h}^{\rm FV} = \overline{d^\prime_{iL}}\,\frac{D^{d}_{il}}{2 v} \,\tilde{\Theta}^R_{lj}\, d^\prime_{jR}\, h +h.c.\,,
\end{eqnarray}
where $\bm{D^d} = \mathrm{diag}(m_d, m_s, m_b)$ is the diagonal mass matrix of the down-type quarks, and with $\bm{\tilde{\Theta}}^R$ defined as $\bm{\tilde{\Theta}}^L$ in Eq. (\ref{eq:ZFV}) after the replacement $L \to R$.

Finally, the axion field, $a$, appears in the phase of the scalar singlet as $\sigma = \frac{v_\sigma + s}{\sqrt{2}}\exp{\left(\frac{i \mathcal{Q}_\sigma a}{v_\sigma}\right)}$, where $\mathcal{Q}_\sigma = 1/2$ is the PQ charge of $\sigma$, as shown in Table \ref{tab:general}. By writing the right-handed quarks $\Psi_{aR}$, which are also charged under $PQ$, as $\Psi_{aR}\to \exp{\left(\frac{i \mathcal{Q}_R a}{ v_\sigma}\right)} \Psi_{aR}$, with $\mathcal{Q}_R = - \mathcal{Q}_\sigma =-1/2$, we can remove the axion field from the Yukawa Lagrangian. As a result, the axion shows up in the kinetic term for $\Psi_{aR}$, which, in turn, mixes with the SM down-type quarks, and after using the Eqs. (\ref{eq:qtrans}) and (\ref{eq:qmix}), we can show that it mediates the flavour-violating interactions below
\begin{eqnarray}\label{eq:FCNCa}
    \mathcal{L}_a^{\rm FV} &=& \frac{ \partial_\mu a}{ v_\sigma} \overline{d^\prime_{iR}}\,\gamma^\mu \,\mathcal{Q}_R\tilde{\Theta}^R_{ij}\, d^\prime_{jR} = \frac{ \partial_\mu a}{f_a} \overline{d^\prime_{i}}\,\gamma^\mu \,\left( c^V_{ij}  + c^A_{ij} \gamma_5 \right)\, d^\prime_{jR}
    \,, \\
    \mbox{with}\quad\quad && c^V_{ij} = c^A_{ij} = c_{ij} =\frac{1}{2}\mathcal{Q}_R\tilde{\Theta}^R_{ij}\,.\nonumber
\end{eqnarray}
Integrating by parts and using the equations of motion for the quarks, we can rewrite it as
\begin{equation}
    \mathcal{L}_a^{\rm FV} = -\frac{i a}{f_a}\left[(m_{d_i}-m_{d_j})\,c_{ij} \,\overline{d^\prime_i}\, d_j^\prime+(m_{d_i}+m_{d_j}) \,c_{ij}\,\overline{d^\prime_i}\,\gamma_5\, d_j^\prime\right]\,. 
\end{equation}

Notice that in contrast with $h$ and $a$, suppressed by $\bm{\tilde{\Theta}}_R$, the leading $Z$-mediated flavour-violating interactions are suppressed by $\bm{\tilde{\Theta}}_L \ll \bm{\tilde{\Theta}}_R$ and, as such, can be safely neglected.

\section{Loop integrals}\label{app:LoopInts}

The integrals in Eq. (\ref{eq:Iints}) are solved following Ref. \cite{AristizabalSierra:2014wal}, and we obtain 
\begin{align}
    \label{eq:loopInt1}
     {\cal I}^{\{2 m_{B^\prime_\alpha} m_{B^\prime_\beta}\}}_{\alpha\beta,ij} &=  \frac{2\,\pi^{4}}{(2\pi)^8} \,\sqrt{r_\alpha\, r_\beta} \,\mathcal{X}_{\alpha\beta,ij}\ ,\\
    \label{eq:loopInt2}
    {\cal I}_{\alpha\beta,ij}^{\{k^2\}} &=\frac{\pi^{4}}{(2\pi)^8}
    \left[
      \frac{ - \hat g(r_\alpha,t_j) 
        + \hat g(r_\alpha,r_\beta)}{t_j-r_\beta}
      +t_i \,\mathcal{X}_{\alpha\beta,ij}
    \right]
    \ ,
    \\
    \label{eq:loopInt3}
    \,{\cal I}_{\alpha\beta,ij}^{\{p^2\}} &= \frac{\pi^{4}}{(2\pi)^8}
        \left[
      \frac{- \hat g(t_i,r_\beta) 
        + \hat g(r_\alpha,r_\beta)}{t_i-r_\alpha}
      + t_j\,\mathcal{X}_{\alpha\beta,ij}
    \right]\ ,
    \\
    \label{eq:loopInt4}
    {\cal I}_{\alpha\beta,ij}^{\{-(p+k)^2\}} &=-\frac{\pi^{4}}{(2\pi)^8}
    \left[
      \widehat B_0^\prime(0,r_\alpha,t_i)\widehat B_0^\prime(0,r_\beta,t_j)+
      \,\mathcal{X}_{\alpha\beta,ij}
    \right]
    \ ,
\end{align}
where
\begin{align}
    \label{eq:loopIntX}
     \mathcal{X}_{\alpha\beta,ij} &=\frac{-{\hat g}(t_i,t_{j})
    +{\hat g}(r_\alpha,t_{j})
    +{\hat g}(t_i,r_\beta)
    -{\hat g}(r_\alpha,r_\beta)}{(t_i-r_\alpha)(t_{j}-r_\beta)}\,,
\end{align}
with $r_\alpha =(m_{B^\prime_\alpha}/m_{\Sigma})^2$ and $t_{i}=(m_{\phi_i}/m_{\Sigma})^2$.
Finally, 
\begin{equation}
  \label{eq:finite-B0}
  \widehat B_0^\prime(0,s,t)=- i\left(\frac{s \log s - t \log t}{s-t}\right)\,\ ,
\end{equation}
comes from to the finite part of the Passarino-Veltmann function $B_0$, and the finite piece for $\hat g(s,t)$ reads:
\begin{eqnarray}\label{eq:ghat}
{\hat g}(s,t) & = & \frac{s}{2}\ln s \ln t +
\sum_{\pm} \pm \frac{s(1-s)+3st+2(1-t)x_{\pm}}{2\omega}
\\ \nonumber
& \times & \Big[\text{Li}_2\left(\frac{x_{\pm}}{x_{\pm}-s}\right) 
              -\text{Li}_2\left(\frac{x_{\pm}-s}{x_{\pm}}\right) 
              +\text{Li}_2\left(\frac{t-1}{x_{\pm}}\right) 
              -\text{Li}_2\left(\frac{t-1}{x_{\pm}-s}\right)  \Big]\ ,
\end{eqnarray}
with the standard di-logarithm
\begin{equation}\label{eq:dilog}
\text{Li}_2(x) = - \int_0^x \frac{\ln(1-y)}{y}dy\ ,
\end{equation}
and
\begin{eqnarray}\label{eq:xpm}
x_{\pm}=\frac{1}{2}(-1+s+t \pm \omega) & & \omega=\sqrt{1+s^2+t^2-2(s+t+st)}\ .
\end{eqnarray}

\end{appendix}
\bibliographystyle{utphys}
\bibliography{bibliography}
\end{document}